\documentclass[ALICE,manyauthors]{cernphprep}
\usepackage[comma,square,numbers,sort&compress]{natbib}

\usepackage{lineno}
\usepackage{xspace}
\usepackage{hyperref}

\usepackage{xcolor}

\usepackage{enumitem} 
\usepackage{dirtytalk} 

\usepackage[T1]{fontenc}
\usepackage{orcidlink}

\begin{document}
%
\newcommand{\lint}         {\ensuremath{\mathcal{L}_\text{int}}}
\newcommand{\kt}{\ensuremath{k_\mathrm{T}}\xspace}
\newcommand{\etajetrange}[1]  {\mbox{$\left | \eta_{\text{jet}} \right |~<~#1$}}
\newcommand{\ptjetch}{\ensuremath{p_\mathrm{T,ch\;jet}}\xspace}
\newcommand{\rg}{\ensuremath{\theta_{\mathrm{g}}}\xspace}
\newcommand{\zg}{\ensuremath{z_\mathrm{g}}\xspace}
\newcommand{\zcut}{\ensuremath{z_{\text{cut}}}\xspace}

\providecommand{\PYTHIA} {{\textsc{pythia 8}}\xspace}
\providecommand{\fastjet} {{\textsc{fastjet}}\xspace}
\providecommand{\GEANT} {{\textsc{geant3}}\xspace}
\providecommand{\HERWIG} {{\textsc{herwig}}\xspace}
\providecommand{\POWHEG} {{\textsc{powheg}}\xspace}
\providecommand{\JEWEL} {{\textsc{jewel}}\xspace}
\providecommand{\JETSCAPE} {{\textsc{jetscape}}\xspace}
\providecommand{\Hybrid} {{\textsc{Hybrid}}\xspace}

\newcommand{\pp}           {pp\xspace}
\newcommand{\ppbar}        {\mbox{$\mathrm {p\overline{p}}$}\xspace}
\newcommand{\XeXe}         {\mbox{Xe--Xe}\xspace}
\newcommand{\PbPb}         {\mbox{Pb--Pb}\xspace}
\newcommand{\pA}           {\mbox{pA}\xspace}
\newcommand{\pPb}          {\mbox{p--Pb}\xspace}
\newcommand{\AuAu}         {\mbox{Au--Au}\xspace}
\newcommand{\dAu}          {\mbox{d--Au}\xspace}

\newcommand{\s}            {\ensuremath{\sqrt{s}}\xspace}
\newcommand{\snn}          {\ensuremath{\sqrt{s_{\mathrm{NN}}}}\xspace}
\newcommand{\pt}           {\ensuremath{p_{\rm T}}\xspace}
\newcommand{\meanpt}       {$\langle p_{\mathrm{T}}\rangle$\xspace}
\newcommand{\ycms}         {\ensuremath{y_{\rm CMS}}\xspace}
\newcommand{\ylab}         {\ensuremath{y_{\rm lab}}\xspace}
\newcommand{\etarange}[1]  {\mbox{$\left | \eta \right |~<~#1$}}
\newcommand{\yrange}[1]    {\mbox{$\left | y \right |~<~#1$}}
\newcommand{\dndy}         {\ensuremath{\mathrm{d}N_\mathrm{ch}/\mathrm{d}y}\xspace}
\newcommand{\dndeta}       {\ensuremath{\mathrm{d}N_\mathrm{ch}/\mathrm{d}\eta}\xspace}
\newcommand{\avdndeta}     {\ensuremath{\langle\dndeta\rangle}\xspace}
\newcommand{\dNdy}         {\ensuremath{\mathrm{d}N_\mathrm{ch}/\mathrm{d}y}\xspace}
\newcommand{\Npart}        {\ensuremath{N_\mathrm{part}}\xspace}
\newcommand{\Ncoll}        {\ensuremath{N_\mathrm{coll}}\xspace}
\newcommand{\dEdx}         {\ensuremath{\textrm{d}E/\textrm{d}x}\xspace}
\newcommand{\RpPb}         {\ensuremath{R_{\rm pPb}}\xspace}

\newcommand{\nineH}        {$\sqrt{s}~=~0.9$~Te\kern-.1emV\xspace}
\newcommand{\seven}        {$\sqrt{s}~=~7$~Te\kern-.1emV\xspace}
\newcommand{\twoH}         {$\sqrt{s}~=~0.2$~Te\kern-.1emV\xspace}
\newcommand{\twosevensix}  {$\sqrt{s}~=~2.76$~Te\kern-.1emV\xspace}
\newcommand{\five}         {$\sqrt{s}~=~5.02$~Te\kern-.1emV\xspace}
\newcommand{\twosevensixnn}{$\sqrt{s_{\mathrm{NN}}}~=~2.76$~Te\kern-.1emV\xspace}
\newcommand{\fivenn}       {$\sqrt{s_{\mathrm{NN}}}~=~5.02$~Te\kern-.1emV\xspace}
\newcommand{\LT}           {L{\'e}vy-Tsallis\xspace}
\newcommand{\GeVc}         {Ge\kern-.1emV/$c$\xspace}
\newcommand{\MeVc}         {Me\kern-.1emV/$c$\xspace}
\newcommand{\TeV}          {Te\kern-.1emV\xspace}
\newcommand{\GeV}          {Ge\kern-.1emV\xspace}
\newcommand{\MeV}          {Me\kern-.1emV\xspace}
\newcommand{\GeVmass}      {Ge\kern-.2emV/$c^2$\xspace}
\newcommand{\MeVmass}      {Me\kern-.2emV/$c^2$\xspace}
\newcommand{\lumi}         {\ensuremath{\mathcal{L}}\xspace}

\newcommand{\ITS}          {\rm{ITS}\xspace}
\newcommand{\TOF}          {\rm{TOF}\xspace}
\newcommand{\ZDC}          {\rm{ZDC}\xspace}
\newcommand{\ZDCs}         {\rm{ZDCs}\xspace}
\newcommand{\ZNA}          {\rm{ZNA}\xspace}
\newcommand{\ZNC}          {\rm{ZNC}\xspace}
\newcommand{\SPD}          {\rm{SPD}\xspace}
\newcommand{\SDD}          {\rm{SDD}\xspace}
\newcommand{\SSD}          {\rm{SSD}\xspace}
\newcommand{\TPC}          {\rm{TPC}\xspace}
\newcommand{\TRD}          {\rm{TRD}\xspace}
\newcommand{\VZERO}        {\rm{V0}\xspace}
\newcommand{\VZEROA}       {\rm{V0A}\xspace}
\newcommand{\VZEROC}       {\rm{V0C}\xspace}
\newcommand{\Vdecay} 	   {\ensuremath{V^{0}}\xspace}

\newcommand{\ee}           {\ensuremath{e^{+}e^{-}}} 
\newcommand{\pip}          {\ensuremath{\pi^{+}}\xspace}
\newcommand{\pim}          {\ensuremath{\pi^{-}}\xspace}
\newcommand{\kap}          {\ensuremath{\rm{K}^{+}}\xspace}
\newcommand{\kam}          {\ensuremath{\rm{K}^{-}}\xspace}
\newcommand{\pbar}         {\ensuremath{\rm\overline{p}}\xspace}
\newcommand{\kzero}        {\ensuremath{{\rm K}^{0}_{\rm{S}}}\xspace}
\newcommand{\lmb}          {\ensuremath{\Lambda}\xspace}
\newcommand{\almb}         {\ensuremath{\overline{\Lambda}}\xspace}
\newcommand{\Om}           {\ensuremath{\Omega^-}\xspace}
\newcommand{\Mo}           {\ensuremath{\overline{\Omega}^+}\xspace}
\newcommand{\X}            {\ensuremath{\Xi^-}\xspace}
\newcommand{\Ix}           {\ensuremath{\overline{\Xi}^+}\xspace}
\newcommand{\Xis}          {\ensuremath{\Xi^{\pm}}\xspace}
\newcommand{\Oms}          {\ensuremath{\Omega^{\pm}}\xspace}
\newcommand{\degree}       {\ensuremath{^{\rm o}}\xspace}

\begin{titlepage}
\PHyear{2026}       
\PHnumber{258}      
\PHdate{10 September}  

\title{Probing jet quenching via the correlation of groomed jet substructure observables in Pb--Pb and pp collisions}
\ShortTitle{Probing jet-quenching via correlation of groomed substructure observables}   

\Collaboration{ALICE Collaboration\thanks{See Appendix~\ref{app:collab} for the list of collaboration members}}
\ShortAuthor{ALICE Collaboration} 
\begin{abstract}
A measurement of the correlation between the splitting angle \rg and the momentum-sharing  fraction \zg of the first hard splitting in a parton shower in \pp collisions and 0--10\% central \PbPb collisions at \fivenn with the ALICE detector is reported.
Charged-particle jets are reconstructed using the anti-\kt algorithm with a jet resolution parameter $R = 0.2$, in the transverse-momentum range $60 \leq \  $\ptjetch $< 80$ \GeVc.
The Soft Drop grooming algorithm is used to identify the first splitting in the parton shower.
Jets observed in \PbPb collisions are narrower than in \pp collisions. This effect is more pronounced for balanced than for unbalanced jets.
No significant modification of the momentum-sharing  fraction is observed for jets with small opening angles.
In contrast, there is a hint that jets with a large opening angle may be less balanced in transverse momentum in \PbPb collisions compared to \pp collisions.
The correlations between \rg and \zg are well described in \pp collisions by \PYTHIA and \POWHEG while \HERWIG shows too few jets with large opening angle. In \PbPb collisions the measurement is compared to a variety of models. The measurement is insensitive to the different implementations of medium response. Inclusion of elastic scatterings, as implemented in the \Hybrid model is preferred to describe the unbalanced jets with large \rg in \PbPb collisions. The balanced jets, however, are less sensitive to elastic scatterings.

\end{abstract}
\end{titlepage}

\setcounter{page}{2} 

\section{Introduction} 
Hadronic matter that reaches a sufficiently high temperature and density will undergo a phase transition into a strongly interacting state of matter known as the quark--gluon plasma (QGP)~\cite{STAR:2005gfr,PHENIX:2004vcz,Muller:2012zq,Braun-Munzinger:2015hba,Busza:2018rrf}.
This plasma can be studied with ultrarelativistic heavy-ion collisions produced at the Relativistic Heavy Ion Collider (RHIC) and the Large Hadron Collider (LHC).

In high-energy particle collisions, partons from the colliding nucleons may undergo interactions with high momentum transfer, leading to highly virtual partons that radiate and form a collimated spray of particles.
These sprays of particles can be experimentally detected, and are known as jets.
In heavy-ion collisions, these jets are formed together with the QGP.
If the particles in the jet interact with the QGP, the properties of the jet can be changed in a process known as \say{jet quenching}.
Measurements of jets in proton--proton (\pp) collisions provide a baseline to isolate the impact of these jet-medium interactions within the QGP~\cite{PhysRevD.27.140,Qin:2015srf,Blaizot:2015lma,Majumder:2010qh}.
Jet quenching has been measured at both RHIC and the LHC for various observables~\cite{STAR:2020xiv,ALICE:2023waz,ATLAS:2018gwx,CMS:2021vui,ALICE:2015mdb,STAR:2017hhs,ATLAS:2022vii,Acharya_2022,CMS:2018fof,ALICE:2017nij,ATLAS:2023hso,CMS:2024zjn}. Jet grooming algorithms aim to select hard-perturbative splittings inside jets, that are calculable from first principles, by removing soft wide-angle radiation from jets~\cite{Larkoski:2014wba,Dasgupta:2013ihk,PhysRevD.91.111501}.
Measurements of groomed jet observables in heavy-ion collisions have opened new possibilities in the study of jet substructure~\cite{CMS:2017qlm,ALICE:2019ykw,CMS:2018fof,ATLAS:2022vii,STAR:2021kjt}.

Soft Drop is a grooming algorithm designed to identify a single, hard splitting within a jet~\cite{Larkoski:2014wba}. 
By removing soft, wide-angle radiation, it reduces the sensitivity of jet measurements to non-perturbative effects such as the underlying event and hadronization. 
This enables more direct comparisons between experimental data and perturbative quantum chromodynamics calculations.
The Soft Drop algorithm is equivalent to the modified mass-drop tagger in the limit of small jet mass~\cite{Dasgupta:2013ihk}.
After reconstructing a jet with a jet algorithm, the jet is reclustered using the Cambridge/Aachen (C/A) algorithm~\cite{Dokshitzer:1997in}.
This will result in an angular ordered shower. The last clustering step in the C/A sequence is reversed, recovering the two subjets from the most recent merging.
If the two subjets of a splitting satisfy the Soft Drop condition, $z > z_{\text{cut}}\theta^\beta$, where $z_{\text{cut}}$ and $\beta$ are tunable parameters, the splitting properties are recorded.
In this expression, $z$ is the fraction of the jet transverse momentum (\pt) carried by the subleading subjet,
\begin{equation}
    z \equiv \frac{p_{\rm{T},\text{subleading}}}{p_{\rm{T},\text{leading}} + p_{\rm{T},\text{subleading}}},
\end{equation}
and $\theta$ is the angular distance between the leading and subleading subjets,
\begin{equation}
    \theta \equiv \frac{\sqrt{\Delta \eta^2+\Delta \varphi^2}}{R},
\end{equation}
where $\Delta \eta$ and $\Delta \varphi$ are the distances in pseudorapidity and azimuthal angle between the subjets, and $R$ is the jet-resolution parameter used in the clustering of the jet. 
The numerator $\sqrt{\Delta \eta^2+\Delta \varphi^2}$ is also commonly known as the (groomed)  splitting angle $R_\mathrm{g}$. If the Soft Drop condition is not met, the lower-energy subjet is discarded.
The declustering procedure is then repeated with the remaining subjet.
If the jet contains no splitting that satisfies the Soft Drop condition, it is removed from the jet sample and recorded for the normalisation.
The percentage of jets in the original jet sample that pass the Soft Drop grooming procedure is quantified by the tagging fraction, $f_{\text{tagged}}$.

Measurements of the groomed observables \zg and \rg have been performed at both RHIC and the LHC in \pp collisions~\cite{ATLAS:2019mgf,CMS:2018ypj,Tripathee:2017ybi,STAR:2020ejj} and heavy-ion collisions~\cite{ALICE:2019ykw,Acharya_2022,CMS:2017qlm,ATLAS:2022vii,CMS:2024zjn}.
The dependence of jet energy loss on the substructure of the jet is expected to result in wider jets losing more energy than narrower jets.
Results from ALICE with a large $z_{\text{cut}} = 0.2$ show no modification of \zg in \PbPb collisions compared to \pp collisions~\cite{Acharya_2022}.
CMS does observe a modification with their measurement at higher jet energy and lower $z_{\text{cut}} = 0.1$~\cite{CMS:2017qlm}.
A selection effect known as survival bias, whereby wider jets — which lose more energy to the medium on average — are less likely to survive the \pt selection threshold, biasing the jet sample toward narrower jets. This survival bias, combined with a rapidly falling jet energy spectrum, will bias any jet sample of a given energy to contain more narrow jets after in-medium jet energy loss.
Measurements of the \rg distribution show a jet narrowing effect in \PbPb collisions compared to \pp collisions, which appears to be independent of jet \pt~\cite{Acharya_2022}. 
This may point to a medium-induced modification of the characteristic splitting angle of groomed jets in \PbPb collisions. However, a measurement of \rg that uses photons as a calibrated probe of the energy scale of the jet shows that survival bias plays an important role in the observed jet narrowing~\cite{CMS:2024zjn}.
Comparing jet narrowing between two samples with similar survival bias may partially disentangle jet shower modifications from survival-bias effects.

Substructure-dependent jet energy loss can be explained by effects such as colour coherence and medium-induced radiation~\cite{Mehtar-Tani:2016aco,Casalderrey-Solana:2012evi}. The pure energy loss that jets suffer through radiation losses, independent of a medium being present or not, can be estimated through the one-dimensional measurements of \rg and \zg. The multi-dimensional approach of the \rg and \zg measurements on the other hand enables detailed investigations of jet quenching at different momentum and angular scales. In case of coherent scatterings, where the angular distance \rg between the leading and subleading jets is smaller than the coherence angle, the \zg distributions should be unaffected. Here the partons remain color correlated and hence emit induced gluons as a single parton. Each of the effective emitters fragments as in vacuum and the total energy loss of the jet is smaller than in the case of incoherent emission off each of its constituent partons~\cite{Casalderrey_Solana_2013}.
For incoherent scatterings, where the decoherence angle is larger than \rg, the \zg should increase as the leading subjet emits energy and \zg should decrease if the subleading subjets loses energy~\cite{Caucal_2021}. 
Only if both subjets suffer from a similar amount of energy loss, the \zg distribution should be insensitive. The angular scale allows to distinguish narrow from wide jets and hence get a handle on the amount of emitters which furthermore scales the energy loss and introduces a survival bias. More narrow jets are expected to survive, in particular if the jet radius is small and aggressive grooming is employed. This can be a limiting factor to study the medium response. 

In this paper, the fully corrected measurement of \rg for selections of large and small \zg in \pp collisions and central \PbPb collisions at \fivenn is reported, together with the fully corrected measurement of \zg for selections of large and small \rg.

\section{Experimental setup and data sets} 
The data for this measurement were recorded with the ALICE detector during LHC Run 2, with \pp and \PbPb collisions collected at a center-of-mass energy of \fivenn using a minimum-bias trigger defined by the coincidence of the signals from two scintillator arrays in the forward region (V0 detectors)~\cite{ALICE-PUBLIC-2018-011,ALICE:2013hur}, and additionally a central trigger for \PbPb collisions.
The central \PbPb collisions are the events with the 10\% highest particle multiplicity, as determined by the forward V0 detectors.
Beam-induced background events and pileup events are rejected~\cite{ALICE:2019qyj}.
The ALICE detector and its performance are described in Refs.~\cite{ALICE2008,ALICE2014,ALICE2022}.
After event selection, the \pp collision data sample contains 870 million events, which corresponds to a total integrated luminosity of $\lint = 18.0 \pm 0.4$ $\text{nb}^{-1}$~\cite{ALICE-PUBLIC-2018-014}.
The central \PbPb collision data sample contains 92 million events, corresponding to an integrated luminosity of $\lint = 0.12$ $\text{nb}^{-1}$~\cite{ALICE-PUBLIC-2021-001}.

The measurements presented in this paper utilize charged particles that are reconstructed with information from the Time Projection Chamber (TPC) and Inner Tracking System (ITS) detectors~\cite{Alme:2010ke,ALICE:2010tia,ALICE:2014sbx}.
In \pp collisions, the Transition Radiation Detector (TRD) is also used for tracking~\cite{ALICE:2017ymw}.
Tracking in the ALICE detector is described in more detail in Ref.~\cite{ALICE-PUBLIC-2020-006}.
Tracks with ${\ensuremath{p_{\rm T,\text{track}}}\xspace} > 0.15 $ \GeVc inside of the pseudorapidity range \etarange0.9 are accepted.
Because of spatial non-uniformity of the ITS response during data taking, different classes of tracks are combined~\cite{ALICE:2012nbx}. 
First, global tracks are required to include at least one hit in the silicon pixel detector (SPD) comprising the first two layers of the ITS and to satisfy commonly used track quality selections.
Then, complementary tracks, i.e. all the tracks that satisfy the selection criteria of global tracks except for the request of a point in the SPD, are refitted using the primary vertex to constrain their trajectory. This preserves a good momentum resolution and an approximately uniform azimuthal acceptance. The momentum resolution of tracks is around 1\% for tracks of $\pt = 1$ \GeVc, and increases to 4\% for tracks at $\pt = 50$ \GeVc.

\section{Analysis method} 
Charged-particle jets are reconstructed using the anti-\kt algorithm in \fastjet 3.2.2~\cite{Matteo_Cacciari_2008,Matteo_Cacciari_2008_2,Cacciari:2011ma}.
The jet resolution parameter is restricted to \textit{R} = 0.2 and the recombination scheme is set to E-scheme, assuming the pion mass for all particles.
Jets are required to have their jet axis pointing within \etajetrange0.7 such that they are fully contained inside the fiducial volume of the TPC.
Any jet that contains a track of ${\ensuremath{p_{\rm T,\text{track}}}\xspace} > 100$ \GeVc is rejected from the jet sample to reduce the influence of fake tracks~\cite{ALICE:2012nbx}.
The measurements of both \pp and \PbPb collision data are corrected for detector effects by using simulated events from the \PYTHIA generator~\cite{Sjostrand:2014zea} with the Monash 2013 tune~\cite{Skands:2014pea}.
All particles in the events obtained from the generator are transported using \GEANT~\cite{Brun:1119728} through a detailed description of the ALICE detector, accounting for the detector response and data-taking conditions. From there the reconstructed-level events are produced.
For the correction of the \PbPb collisions measurement, the reconstructed-level events are embedded into real central \PbPb event data to serve as the underlying background event that is present in heavy-ion collisions. A six-dimensional response matrix is constructed from the matched generated-level and reconstructed-level jets that characterizes the impact of the detector and background effects as function of the transverse momentum of the charged-particle jet (\ptjetch), \zg and \rg.
It also accounts for the tagging fraction of the grooming procedure.
The three-dimensional unfolded results are then obtained by iterative Bayesian unfolding~\cite{DAGOSTINI1995487} using RooUnfold~\cite{RooUnfold}.

The large uncorrelated background present in heavy-ion collisions because of the underlying event is subtracted using the constituent subtraction method~\cite{Berta:2014eza,Berta:2019hnj}.
The event-wide constituent subtraction method applies a correction based on the average background \pt and transverse mass of each event. It is used with a default maximum recombination distance $R_{\text{max}} = 0.1$ and the \pt weighting factor $\alpha = 0$.
This method suppresses background contributions to both the jet momentum and the jet substructure by subtracting the average underlying event at the constituent level.
After background subtraction, jets are required to have $40 \leq \ $\ptjetch $< 120$ \GeVc.
The constituent subtraction method cannot correct for local fluctuations of the
background.
These local fluctuations can influence which splitting is selected by the grooming procedure and can lead to mistags, where the selected splitting is unrelated to the parton shower~\cite{Mulligan:2020tim,Acharya_2022}.
This background may also modify the angular ordering of the shower or deposit additional energy in one of the branches of the jet, resulting in a background-induced splitting being selected in place of the true hard splitting.
To suppress the sensitivity to soft wide-angle radiation that causes falsely tagged splittings grooming is employed with a strong grooming condition with $\zcut = 0.2$ and $\beta = 0$.
This choice of \zcut suppresses the contamination of falsely tagged splittings to a level that permits unfolding for detector and background effects~\cite{Acharya_2022}. Through the parameter $\beta = 0$ the angular distance between the leading and subleading subjets is irrelevant. 
With these grooming conditions, the procedure correctly identifies more than 80\% of subleading prongs in central \PbPb collisions. A detailed study of falsely tagged splittings in ALICE shows that the residual background contribution ranges from approximately 5\% up to 15\% at lower \pt~\cite{ALICE-PUBLIC-2020-006, ALICE:2024fip} .

The \pp collision measurements are compared to predictions from the following event generators.

\begin{itemize}[itemsep=1pt, topsep=1pt]
    \item \PYTHIA is a general-purpose Monte Carlo generator to simulate \pp events~\cite{Sjostrand:2014zea}. 
    \PYTHIA includes hard process matrix elements at leading order.
    The Monash 2013 tune that is tuned to LHC data~\cite{Skands:2014pea} is used.
    \item \POWHEG generates next-to-leading order hard scatterings which are then given to \PYTHIA to perform the parton shower evolution~\cite{Nason:2004rx,Frixione:2007vw,Alioli:2010xd}. The default dijet tune is used.
    \item \HERWIG 7.2.2 is a general-purpose Monte Carlo event generator~\cite{Bahr:2008pv,Bellm:2015jjp}. It distinguishes itself from \PYTHIA because of the angular-ordered shower evolution. The default leading-order LHC tune is used.
\end{itemize}
The measurements in \PbPb collisions are compared with \JEWEL, \JETSCAPE and the \Hybrid model event generators, where the respective baseline \pp predictions are also used when comparing the \pp and \PbPb collision measurements.
\begin{itemize}[itemsep=1pt, topsep=1pt]
    \item \JEWEL is a dedicated jet-quenching Monte Carlo event generator that also comes with a baseline model (\JEWEL vacuum) without jet quenching. The jet-quenching model is based on BDMPS medium-induced gluon radiation in a medium model with Bjorken expansion~\cite{Baier_1997}. The quenching model can include or exclude medium recoils~\cite{Zapp:2013vla}. If medium recoils are included, the uncorrelated final-state particles introduced by the parton-medium interaction are subtracted using the constituent subtraction method~\cite{milhano2022improvedbackgroundsubtractionfresh}.
    \item \JETSCAPE is a modular framework designed to simulate heavy-ion collisions using third-party modules by factorizing physics components. The vacuum sample is generated with vacuum MATTER of the \JETSCAPE PP19 tune~\cite{JETSCAPE:2019udz}, while the \PbPb sample is generated with MATTER+LBT using the \JETSCAPE v3.5 AA22 tune~\cite{JETSCAPE:2022jer, JETSCAPE:2022hcb, JETSCAPE:2023hqn}.
    \item The \Hybrid model~\cite{Casalderrey-Solana:2014bpa,DEramo:2018eoy,Hulcher:2017cpt} is a jet-quenching model that calculates in-medium parton energy loss in a strongly coupled, AdS-CFT inspired manner. It includes a model of medium response in the form of a medium wake — the collective hydrodynamic response of the QGP to the momentum deposited by the jet. In the \Hybrid model, the effect of elastic scatterings~\cite{hulcher2026sensitivityjetobservablesmoliere} and the wake can be independently turned on and off. The \Hybrid model without elastic scatterings and wake is referred to as the \say{default}.
\end{itemize}

\section{Systematic uncertainties}
There are various sources of systematic uncertainty in this measurement which are all addressed by taking the difference between the nominal result and the result obtained by varying each source of systematic uncertainty. For the measurement in \pp collisions, the sources of systematic uncertainty originate from the uncertainty on the tracking efficiency, the unfolding procedure, and the model dependence of the detector response.
In \PbPb collisions, the same sources of systematic uncertainty are present, and additionally the background subtraction procedure induces an uncertainty. 

The large background that is present in \PbPb collisions can influence which splitting is selected during the grooming procedure~\cite{Mulligan_2020}. Residual background effects not captured by the response matrix can cause the unfolding to not recover the true distribution, referred to as non-closure. To assess this, a non-closure systematic uncertainty is estimated through repeating the entire analysis on embedded Monte Carlo data~\cite{ALICE-PUBLIC-2020-006, Acharya_2022}. For this, \PYTHIA events are embedded into a simulated thermal background as is expected to emerge from the QGP. The simulation samples a normal distribution to mimic the charged background particles that are measured within the ALICE acceptance. Then the embedded results are unfolded and compared to the baseline to obtain the systematic uncertainty which is added in quadrature to the total systematic uncertainty of the \PbPb collision measurement.

The systematic uncertainty on the measurement due to the uncertainty on the tracking efficiency is determined by randomly rejecting tracks at the reconstructed level before jet reconstruction.
To assess the track reconstruction efficiency uncertainty, 3\% of the tracks in \pp collisions and 4\% of tracks in \PbPb collisions are randomly rejected. This choice is based on variations in the track selection criteria and on the ITS–TPC track-matching efficiency uncertainty.

The systematic uncertainty of the measurement due to the unfolding procedure is determined by varying the unfolding parameters, which allows to analyse the influence of these parameters on the final result.
The first variation is on the number of iterations of the unfolding procedure, known as the regularization parameter, which is varied by $\pm 2$.
The larger of the two variations is evaluated bin-by-bin and is taken as the final regularization uncertainty.
To assess the impact of binning on the reconstructed level of the response matrix, the number of bins is doubled.
The lower limit of \ptjetch that enters into the response on the reconstructed level is varied by $\pm 10$ \GeVc, and the per bin maximum associated uncertainty of the two variations is used as the final uncertainty.
Finally, the generator level \ptjetch, \rg, and \zg prior distributions obtained from \PYTHIA are reweighted to spectra obtained from \HERWIG~\cite{Bahr:2008pv,Bellm:2015jjp} for \pp collisions and \JEWEL without recoils~\cite{Zapp:2013vla} for \PbPb collisions. The unfolding is repeated with the reweighted priors, and the difference between the unfolded results and the nominal one is taken as the systematic uncertainty from prior shape dependence.
Because all these variations probe the same underlying unfolding systematic uncertainty, the total per bin systematic uncertainty associated with the unfolding procedure is then calculated as the average of the individual variations.

To unfold the measurement for detector effects, simulated events that are propagated through a detailed model of the ALICE detector using GEANT3 are analysed.
The response matrix depends on the generator used to construct it.
This uncertainty is evaluated by comparing \PYTHIA to \HERWIG for \pp collisions, and to \JEWEL without recoils for \PbPb collisions. Any difference observed in the final unfolded result is assigned as a systematic uncertainty from the model dependence.

Underlying-event subtraction via the constituent subtraction method to subtract the uncorrelated particles in \PbPb collisions gives an additional systematic uncertainty that is estimated by performing a relative over- and undersubtraction with respect to the nominal subtraction.
The nominal value of the maximum recombination distance $R_{\text{max}} = 0.1$ is varied to $R_{\text{max}} = 0.05$ and $R_{\text{max}} = 0.5$ and the per-bin maximum difference is taken as the systematic uncertainty in each interval.

Representative values of systematic uncertainties in \pp and \PbPb collisions are given in Tab.~\ref{table:ppSysTable} and Tab.~\ref{table:PbPbSysTable}, respectively.
The final systematic uncertainty for the \pp and \PbPb collision measurements is then calculated as the quadratic sum of uncertainties from all the relevant individual sources.

\begin{table}[h!]
\centering
\begin{tabular}{llll}
\hline
\pp collisions & \multicolumn{3}{l}{Relative uncertainty (Min. -- Max. \%)}                         \\ \cline{2-4} 
                & Trk. eff. & Unfolding & Model dep.  \\ \hline
\rg low \zg       & 2--4\%     & 1--4\%     & 0--1\%                     \\
\rg high \zg      & 1--4\%     & 1--8\%     & 1--4\%                     \\ \hline
\zg low \rg       & 0--2\%     & 2--3\%     & 0--1\%                     \\
\zg high \rg      & 1--4\%     & 2--4\%     & 0--2\%                     \\ \hline
\end{tabular}
\caption{Summary of the sources of systematic uncertainty in the \pp collision measurements. The range corresponds to the minimum and maximum per bin uncertainty due to that source.\label{table:ppSysTable}}
\end{table}

\begin{table}[h!]
\centering
\begin{tabular}{llllll}
\hline
\PbPb collisions & \multicolumn{5}{l}{Relative uncertainty (Min. -- Max. \%)}                         \\ \cline{2-6} 
                & Trk. eff. & Unfolding & Model dep. & Bkgd. sub. & Thermal non-closure \\ \hline
\rg low \zg       & 0--4\%     & 1--5\%     & 0--8\%      & 0--8\%      & 0--4\%               \\
\rg high \zg      & 1--3\%     & 1--4\%     & 1--4\%      & 0--9\%      & 0--1\%               \\ \hline
\zg low \rg       & 0--1\%     & 1--2\%     & 0--2\%      & 1--4\%      & 0--2\%               \\
\zg high \rg      & 2--4\%     & 1--4\%     & 1--3\%      & 2--4\%      & 0--1\%               \\ \hline
\end{tabular}
\caption{Summary of the sources of systematic uncertainty in the \PbPb collision measurements. The range corresponds to the minimum and maximum per bin uncertainty due to that source.\label{table:PbPbSysTable}}
\end{table}

\section{Results}
The two-dimensional correlation of \rg and \zg is reported.
The projected one-dimensional \rg distribution for selections of small and large values of \zg, and vice versa, are extracted from this two-dimensional correlation.
All distributions are normalized by the total number of jets with $60 \le \ptjetch < 80$ \GeVc. Since not all jets satisfy the Soft Drop criterion, the distributions do not integrate to unity but to the tagging fraction $f_{\text{tagged}}$, representing the fraction of jets with an identified splitting. The tagging fraction in \pp and \PbPb collisions is $89\%$ and $87\%$, respectively.
 
\begin{figure}[!h]
\centering
  \includegraphics[width=.46\textwidth]{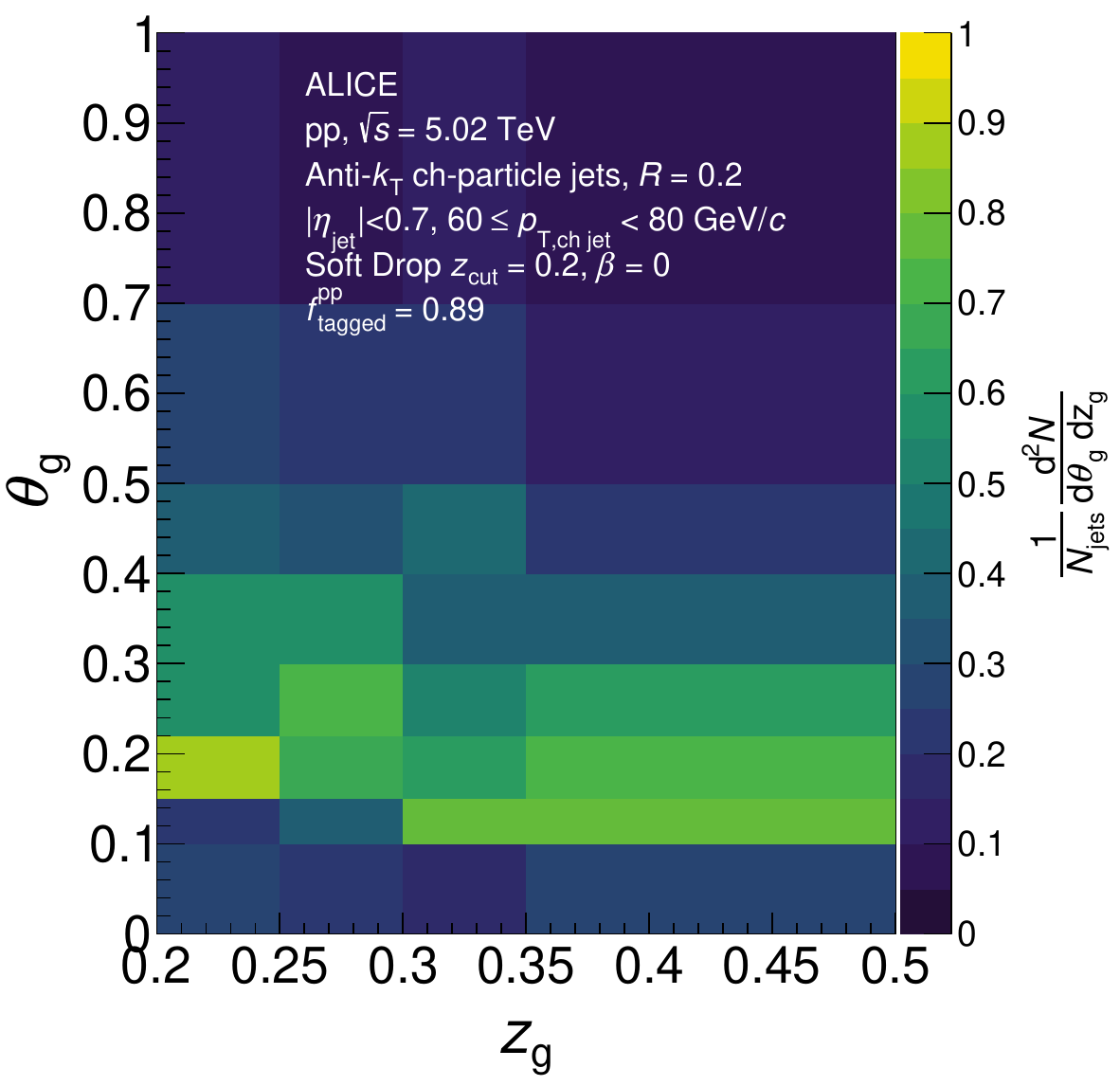}\hfill
  \includegraphics[width=.46\textwidth]{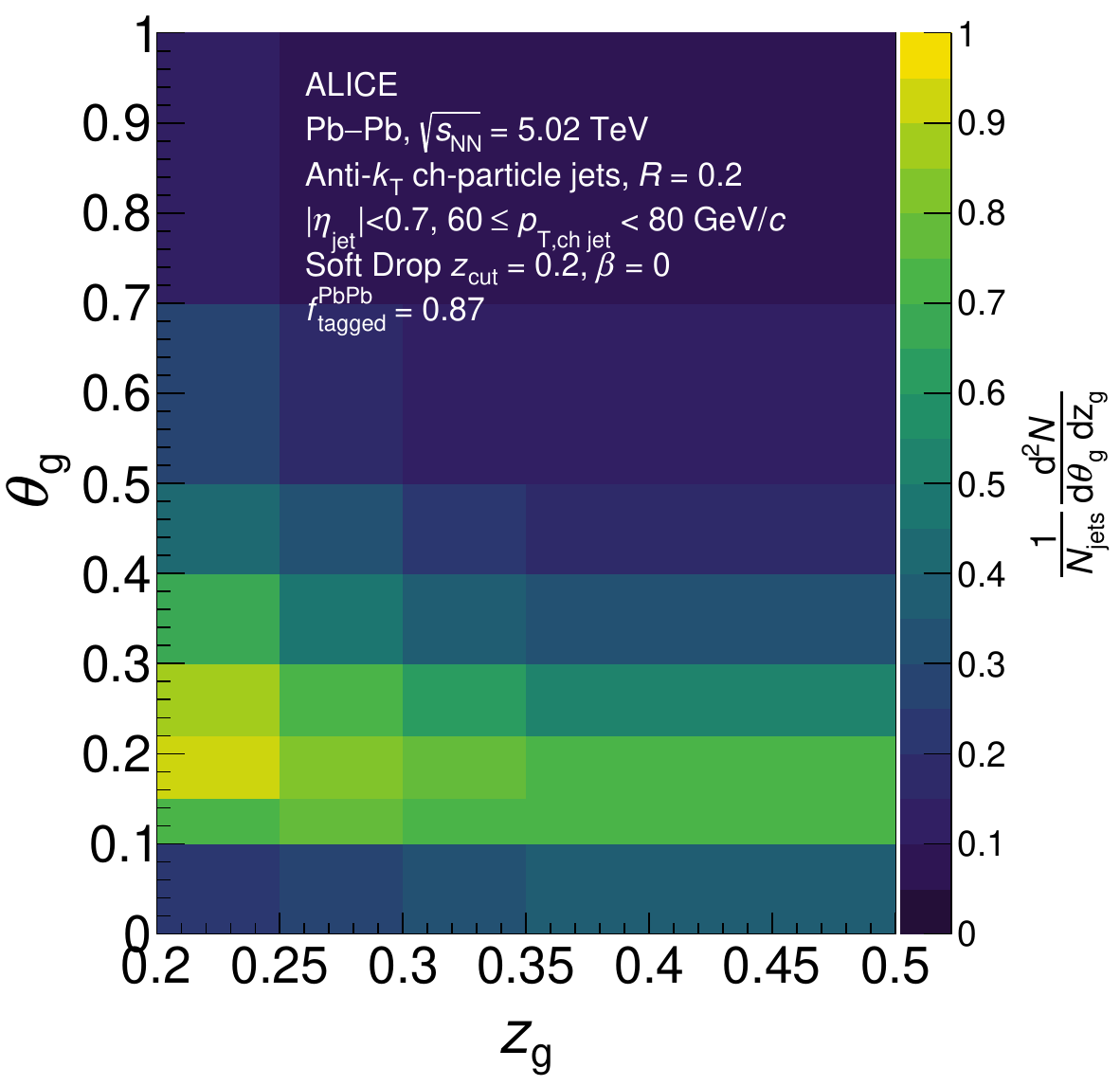}\hfill
  \caption{The two-dimensional correlation of \rg and \zg for \pp (left) and \PbPb (right) collisions at \fivenn for jets with $60 \leq $ \ptjetch $< 80$ \GeVc.}
  \label{fig:DataCorrelations}
\end{figure}

Figure~\ref{fig:DataCorrelations} shows the two-dimensional correlation of \rg and \zg for \pp (left) and \PbPb (right) collisions for jets with transverse momentum $60 \le \ptjetch < 80$ \GeVc.
This fully corrected two-dimensional result represents the jet-by-jet correlation of \rg and \zg.
This result is used to scan the angular and momentum scales of splittings in jets by constructing projections of \rg in two bins of \zg and vice versa.

\begin{figure}[!h]
\centering
  \includegraphics[width=.5\textwidth]{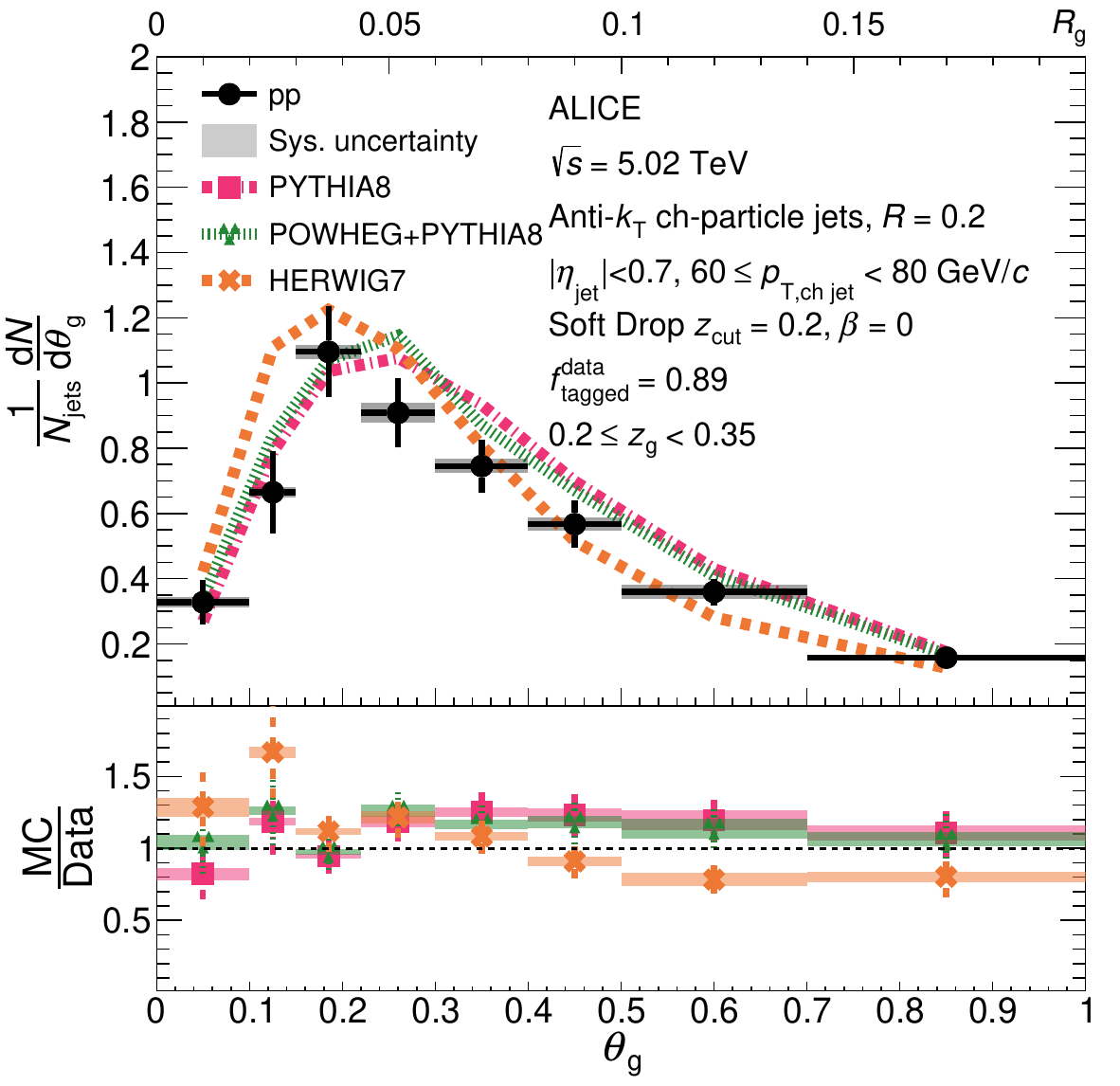}\hfill
  \includegraphics[width=.5\textwidth]{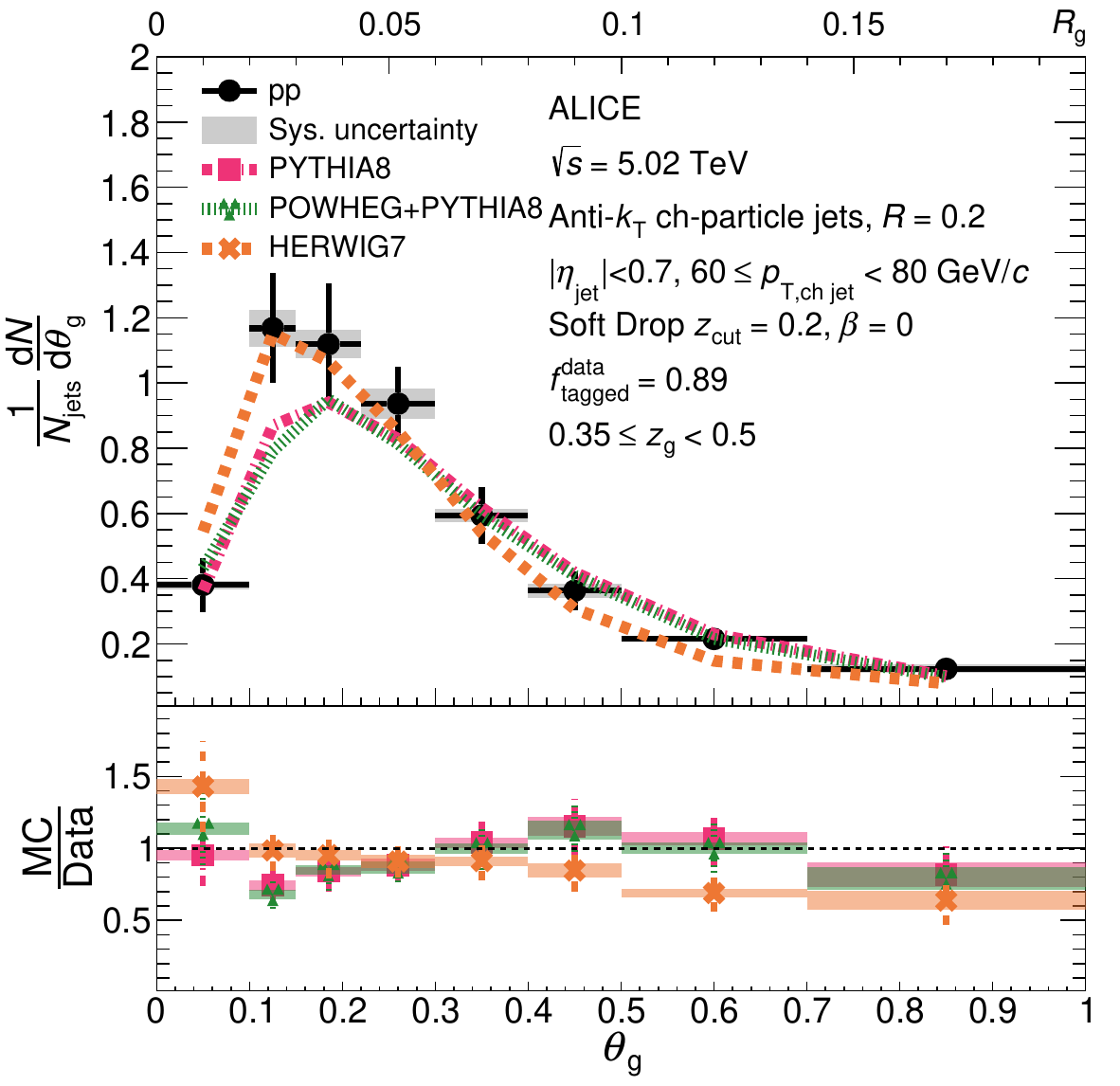}\hfill
  \caption{The distributions of \rg in \pp collisions at \five, for jets with $60 \leq $ \ptjetch $< 80$ \GeVc, for selections of small and large \zg. The left panel shows the \rg distribution for $0.2 \leq \zg < 0.35$, and the right panel shows \rg for $0.35 \leq \zg < 0.5$.}
  \label{fig:ppRgResults}
\end{figure}

\begin{figure}[!h]
\centering
  \includegraphics[width=.5\textwidth]{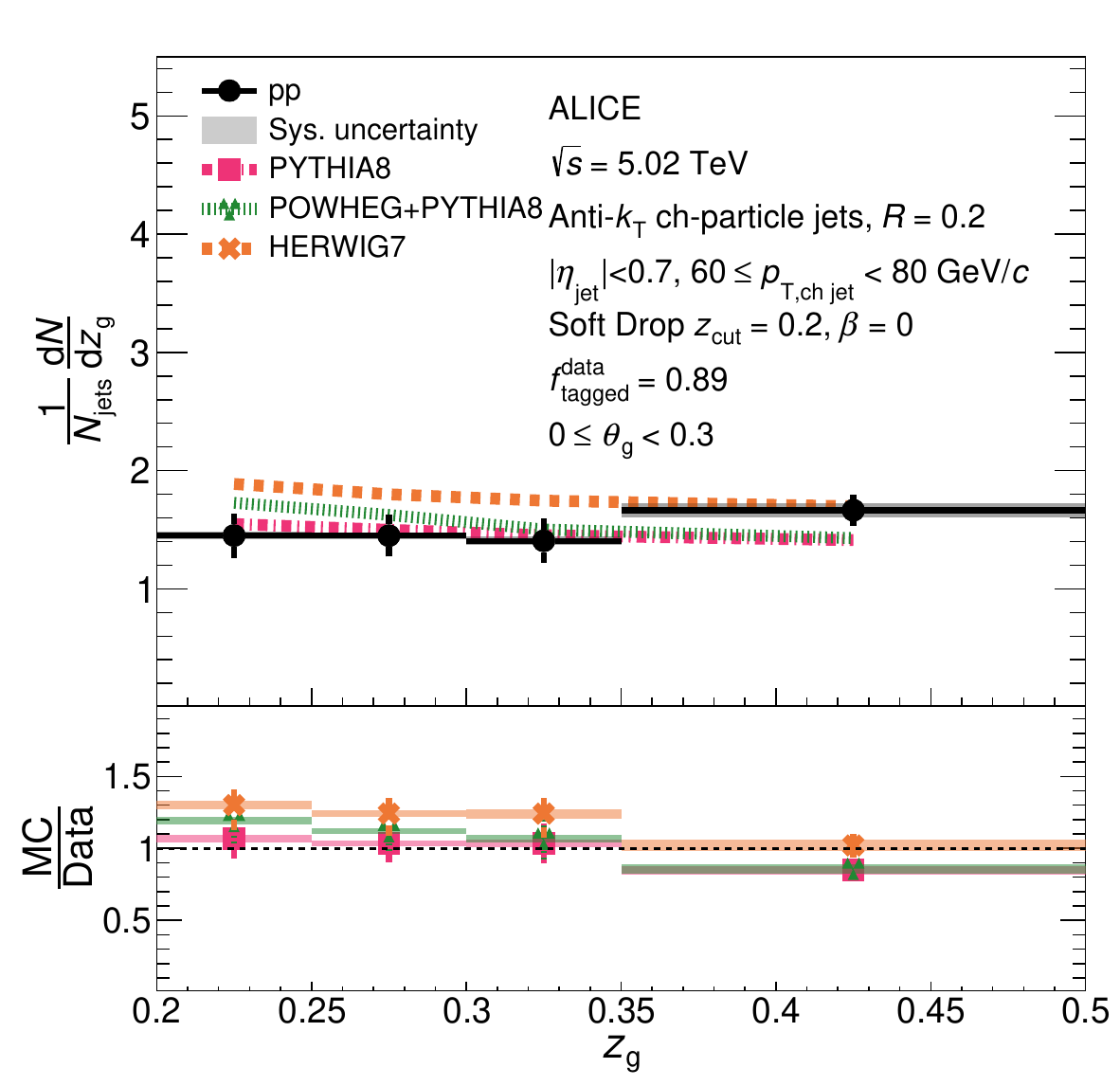}\hfill
  \includegraphics[width=.5\textwidth]{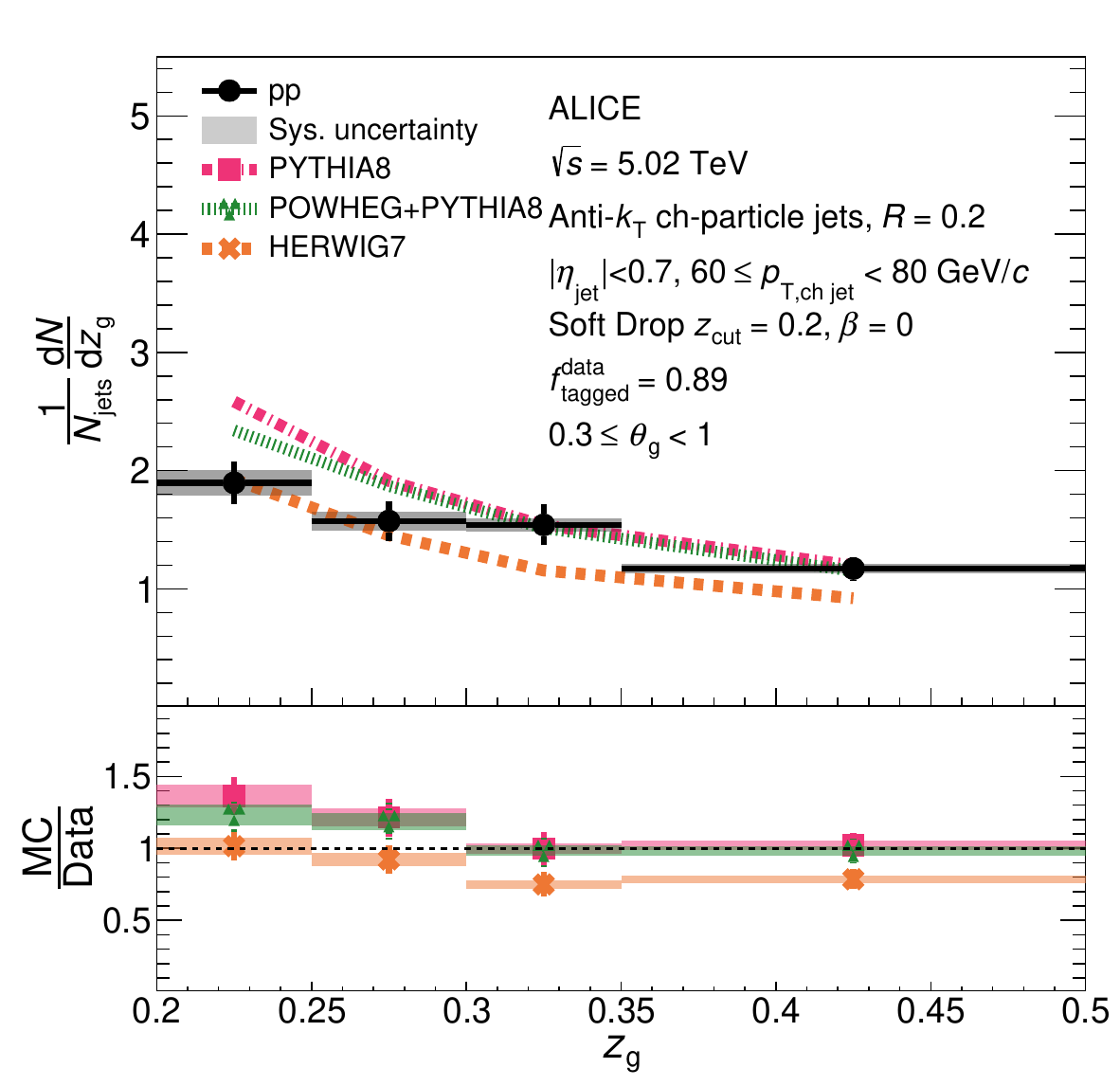}\hfill
  \caption{The distributions of \zg in \pp collisions at \five, for jets with $60 \leq $ \ptjetch $< 80$ \GeVc, for selections of small and large \rg. The left panel shows the \zg distribution for $0.0 \leq \rg < 0.3$, and the right panel shows \zg for $0.3 \leq \rg < 1.0$.}
  \label{fig:ppZgResults}
\end{figure}

Figure~\ref{fig:ppRgResults} shows the distribution of \rg in \pp collisions for selections of small and large \zg, corresponding to more momentum-unbalanced and balanced jets, respectively.
The selection for small \zg contains all jets with  $0.2 \leq \zg < 0.35$, while the jets with large \zg satisfy $0.35 \leq \zg < 0.5$. Balanced subjets, corresponding to large \zg, have on average a smaller splitting angle than unbalanced subjets.
\PYTHIA and \POWHEG agree reasonably well with the \pp data, while for \HERWIG some deviations are observed because \HERWIG produces slightly more narrow and less balanced jets.
Similarly, Fig.~\ref{fig:ppZgResults} shows the \zg distribution in \pp collisions for selections of small and large \rg, which corresponds to narrow and wide jets, respectively.
The selection for small \rg contains all jets with  $0.0 \leq \rg < 0.3$, while the jets with large \rg have $0.3 \leq \rg < 1.0$.
\PYTHIA, \POWHEG, and \HERWIG  agree fairly well with the \pp data.
The \rg and \zg model predictions in \pp collisions for the jet-quenching event generators, which are based on \PYTHIA, are shown in the Appendix in Figures~\ref{fig:baselineRgResults} and~\ref{fig:baselineZgResults}.

\begin{figure}[!h]
\centering
  \includegraphics[width=.5\textwidth]{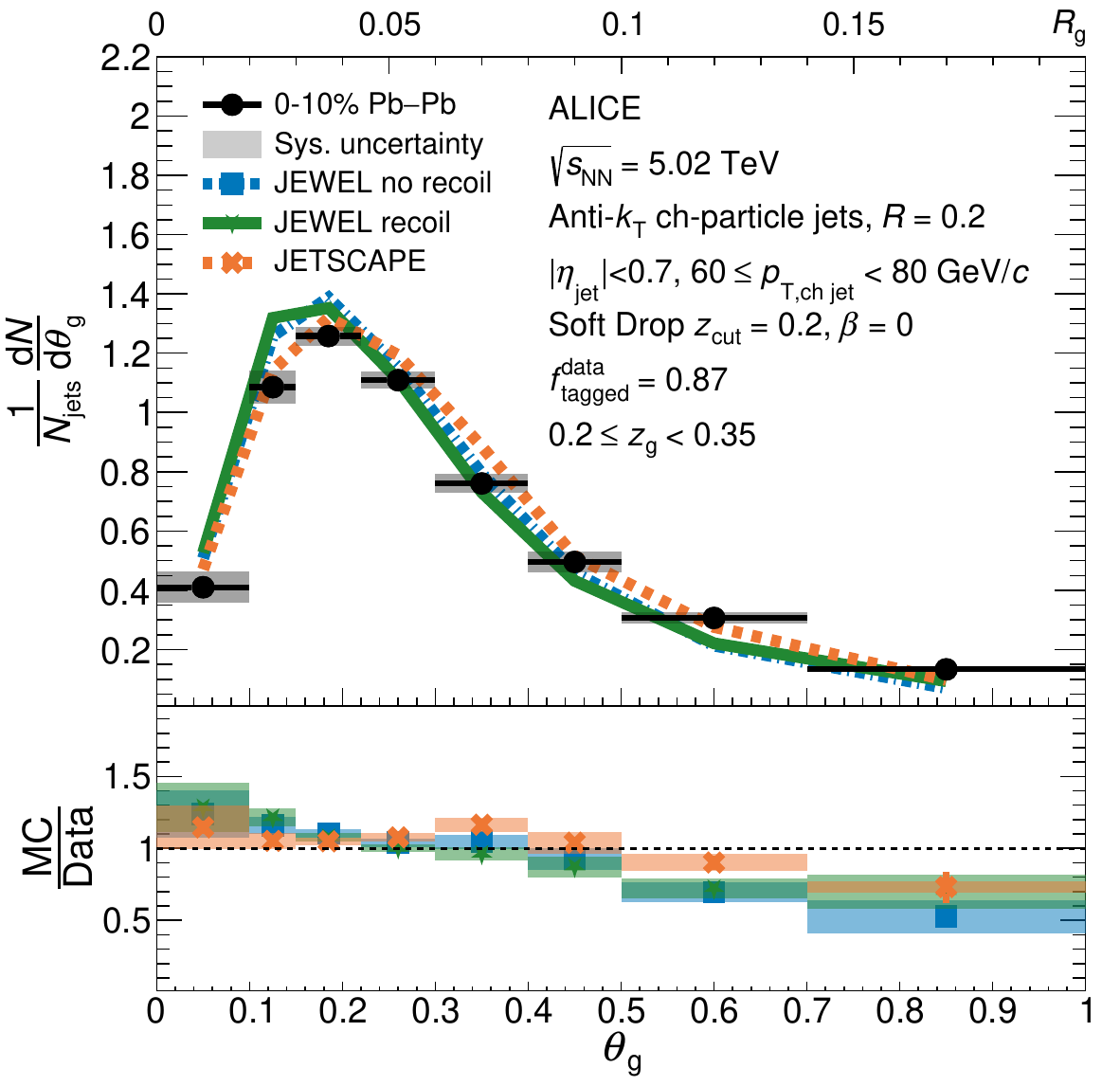}\hfill
  \includegraphics[width=.5\textwidth]{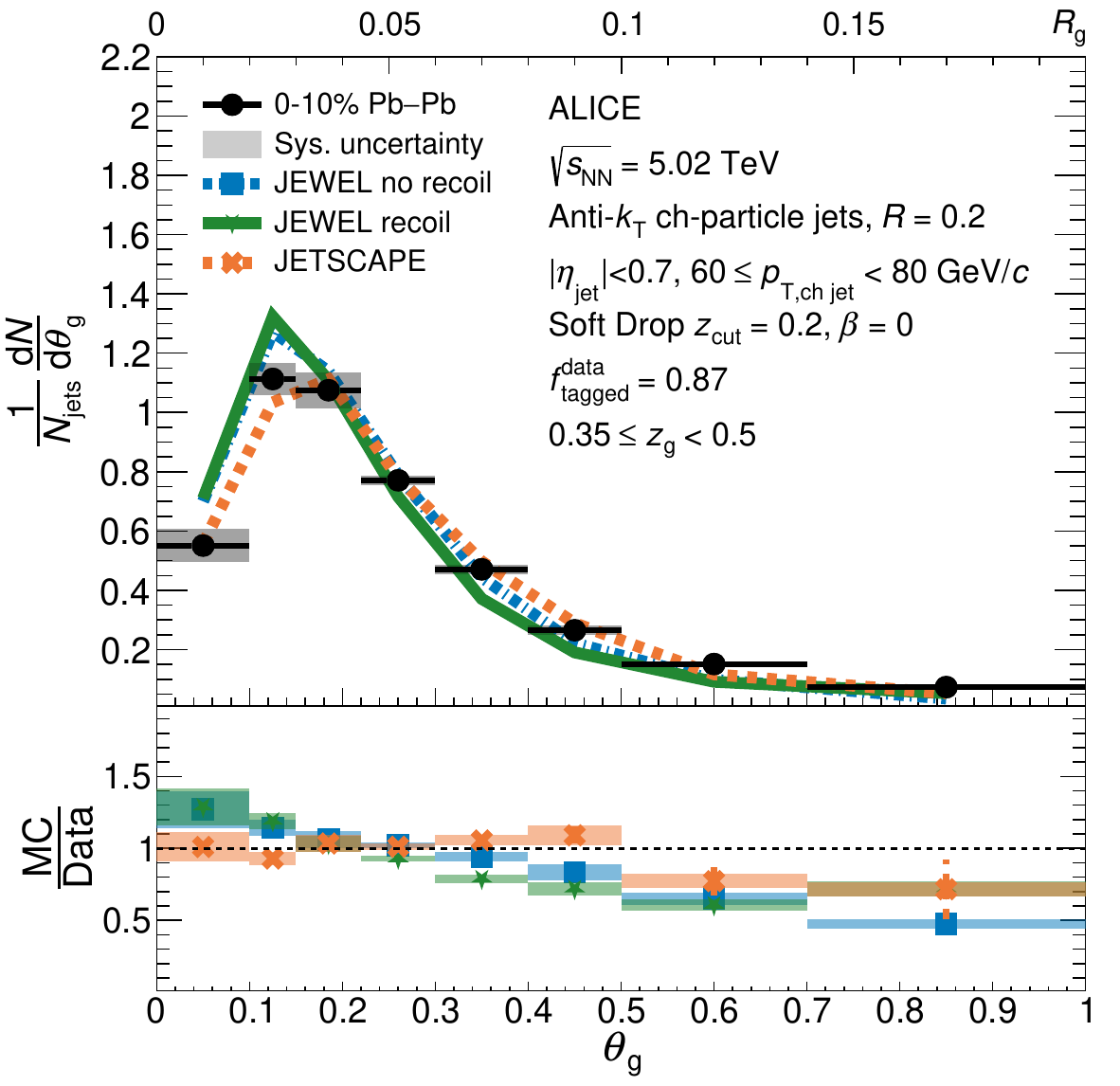}\hfill
  \includegraphics[width=.5\textwidth]{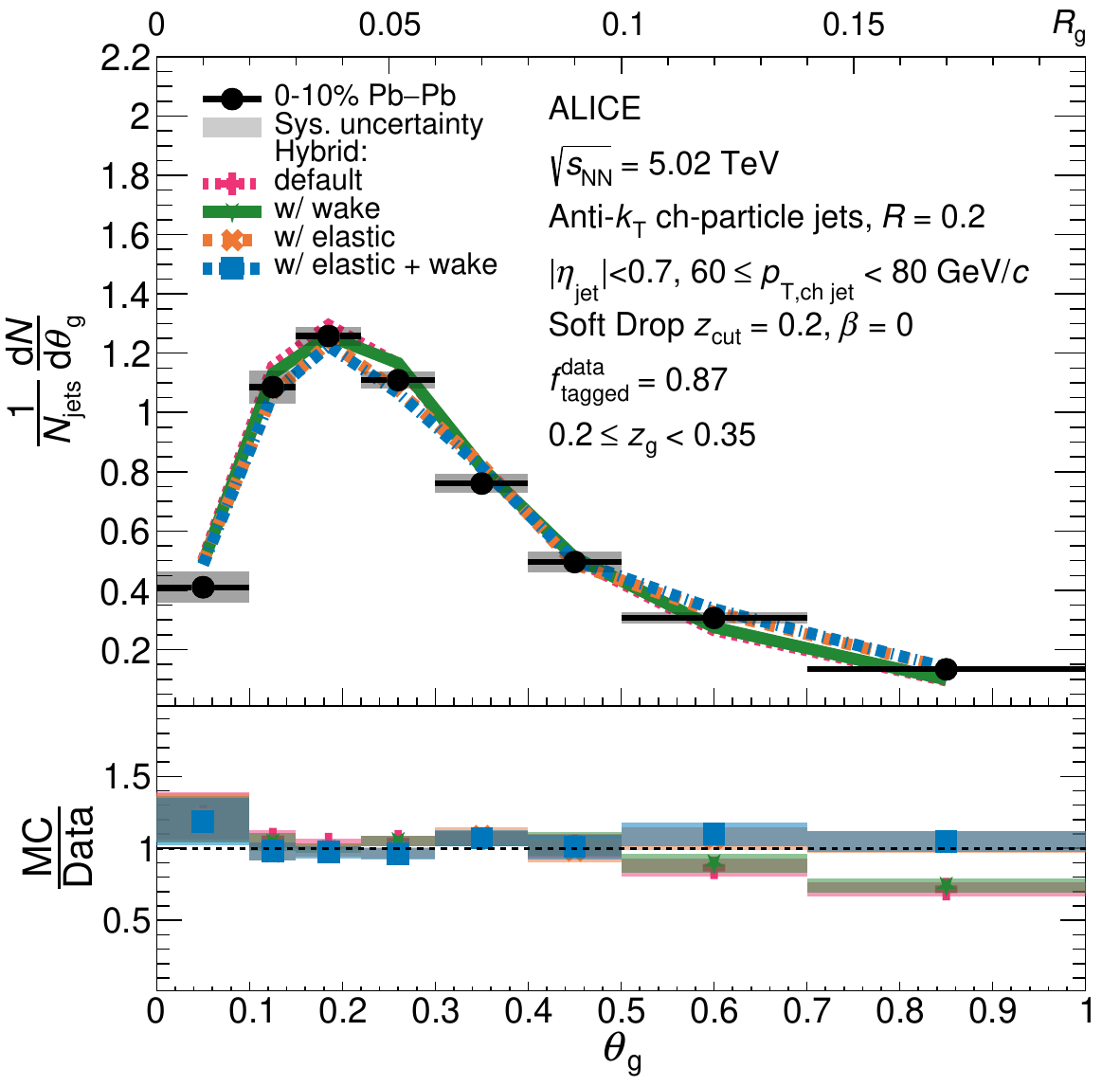}\hfill
  \includegraphics[width=.5\textwidth]{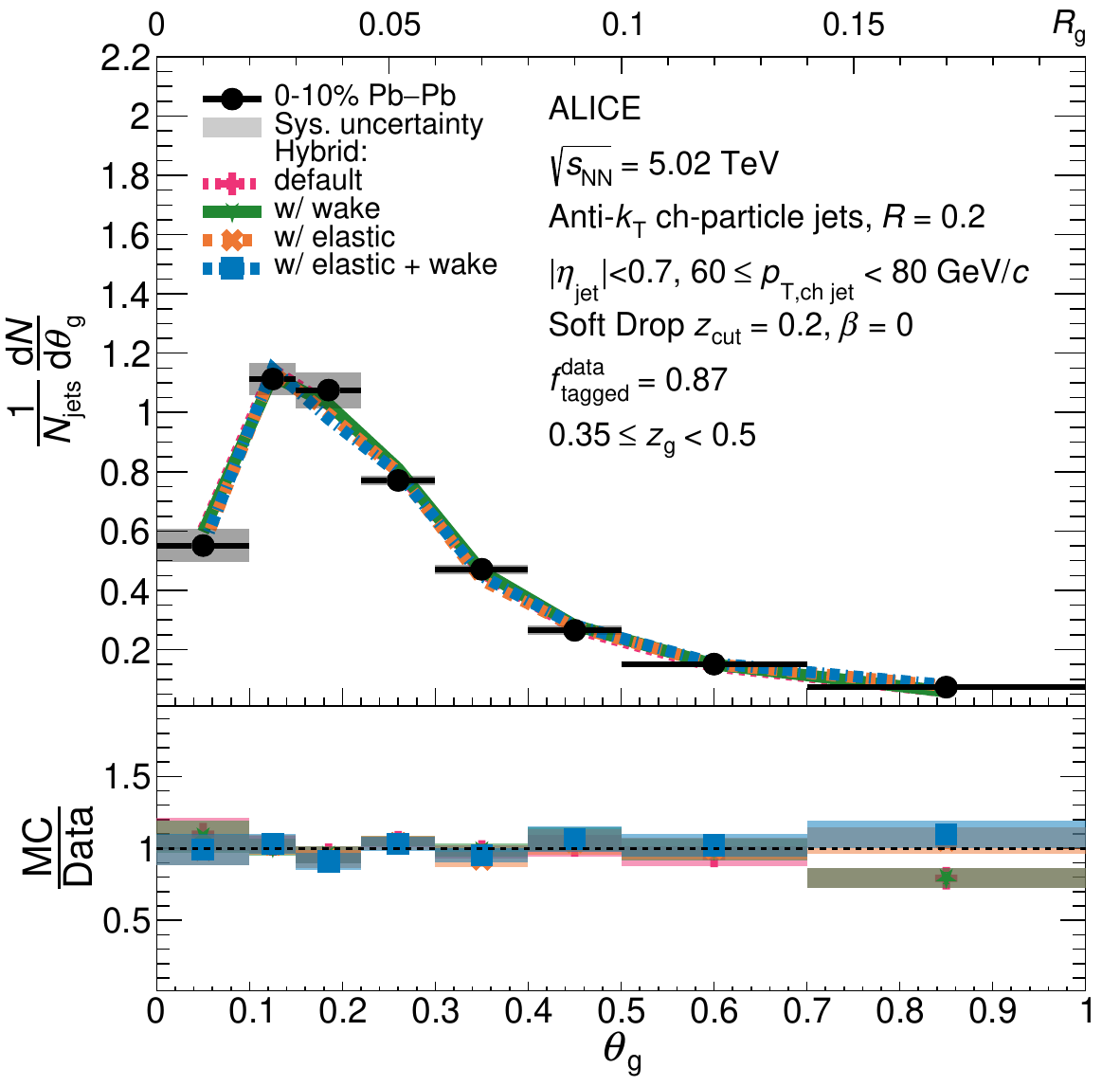}\hfill  
  \caption{The distributions of \rg in \PbPb collisions at \fivenn for jets with $60 \leq $ \ptjetch $< 80$ \GeVc for selections of small and large \zg. The left panels show the \rg distribution for $0.2 \leq \zg < 0.35$, and the right panels show \rg for $0.35 \leq \zg < 0.5$. Top panels compare the \JEWEL and \JETSCAPE models and bottom panels present the different \Hybrid tunes.}
  \label{fig:PbPbRgResults}
\end{figure}

Figure~\ref{fig:PbPbRgResults} shows the distribution of \rg in \PbPb collisions for selections of unbalanced and balanced subjets.
The selection in \zg is the same as for \pp collisions.
The \PbPb collision data is compared to \JEWEL, \JETSCAPE and the \Hybrid model.
\JEWEL overestimates the data in the \rg distribution at small angles and underestimates it at large angles for both selections in \zg, regardless of whether recoils are included or not. 
At the last \rg interval, where the effect of recoil is most prominent in the predictions, the data favours recoil over no recoil.
\JETSCAPE describes the small \rg region of both unbalanced and balanced subjets well, but there is a hint of underestimation of the large \rg contribution for both selections.
The \Hybrid model predictions of \rg show that this measurement, where aggressive grooming is applied to jets of small radii, is fairly insensitive to the effects of the medium wake. 
However, elastic scatterings seem to have an impact in the large \rg region that is slightly more pronounced for unbalanced jets. For both, the unbalanced and balanced subjet populations, the data is more consistent with the \Hybrid prediction that includes elastic scatterings and possibly wake effects over the one without elastic scatterings.

\begin{figure}[!h]
\centering
  \includegraphics[width=.5\textwidth]{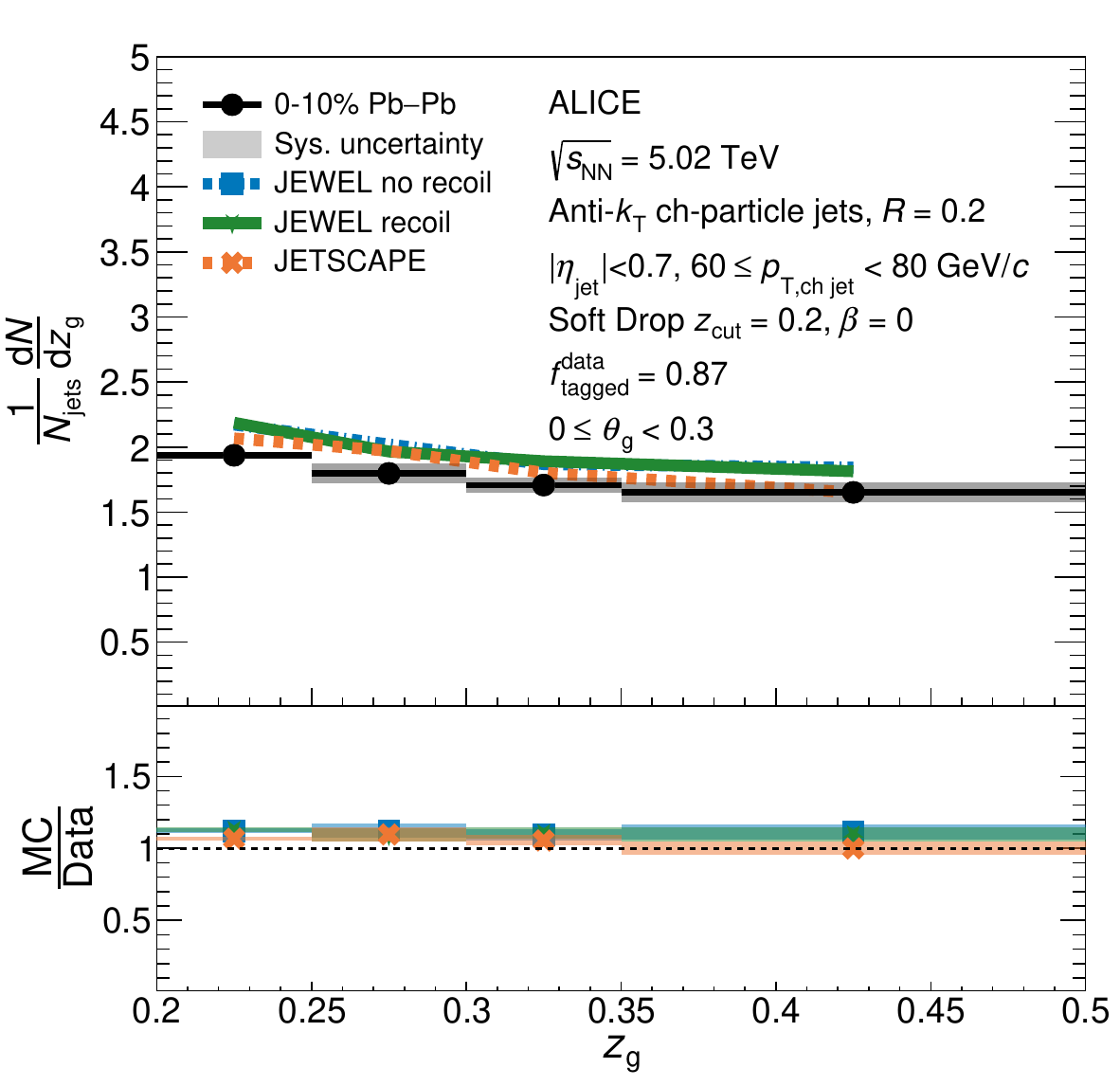}\hfill
  \includegraphics[width=.5\textwidth]{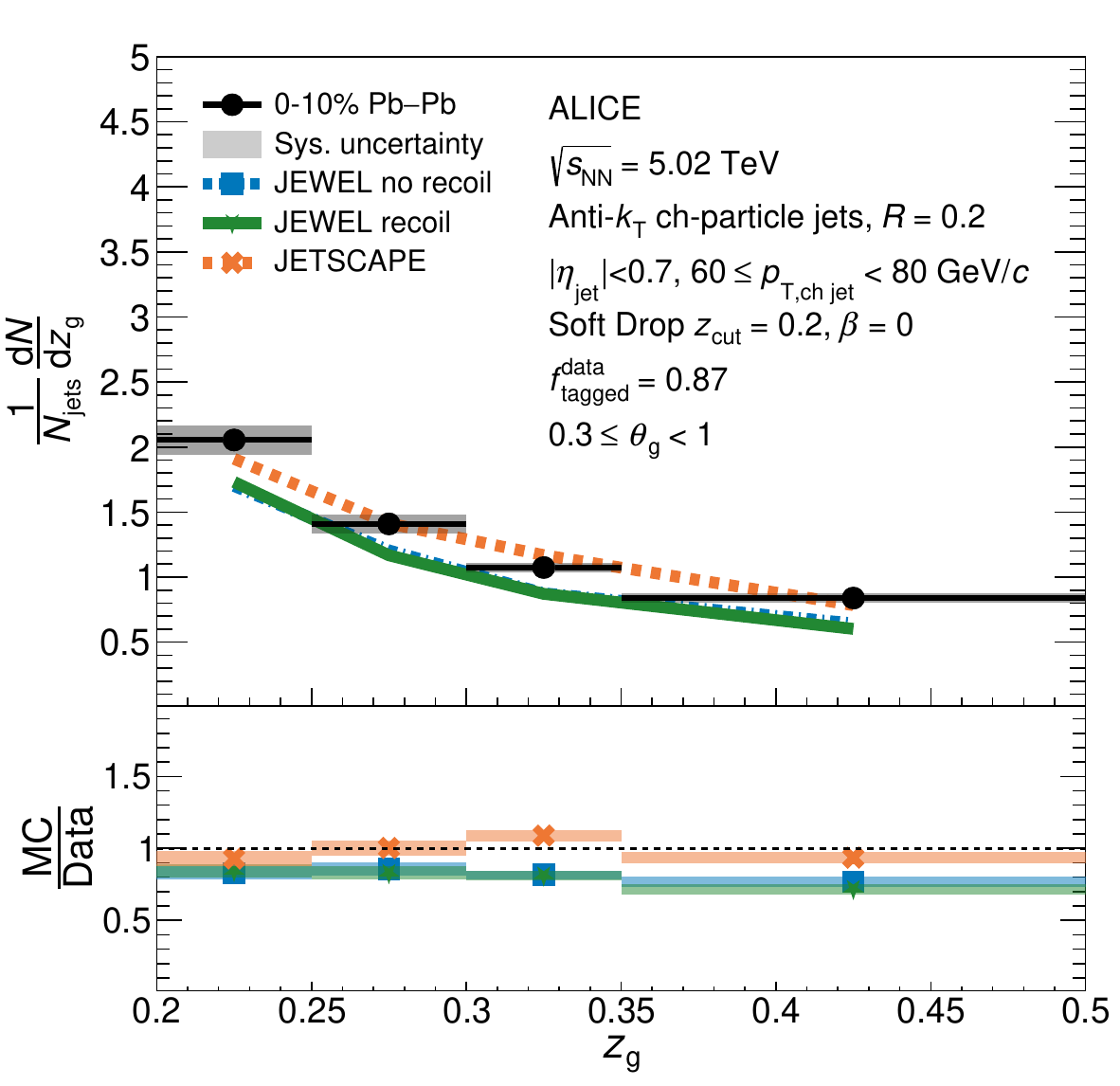}\hfill
  \includegraphics[width=.5\textwidth]{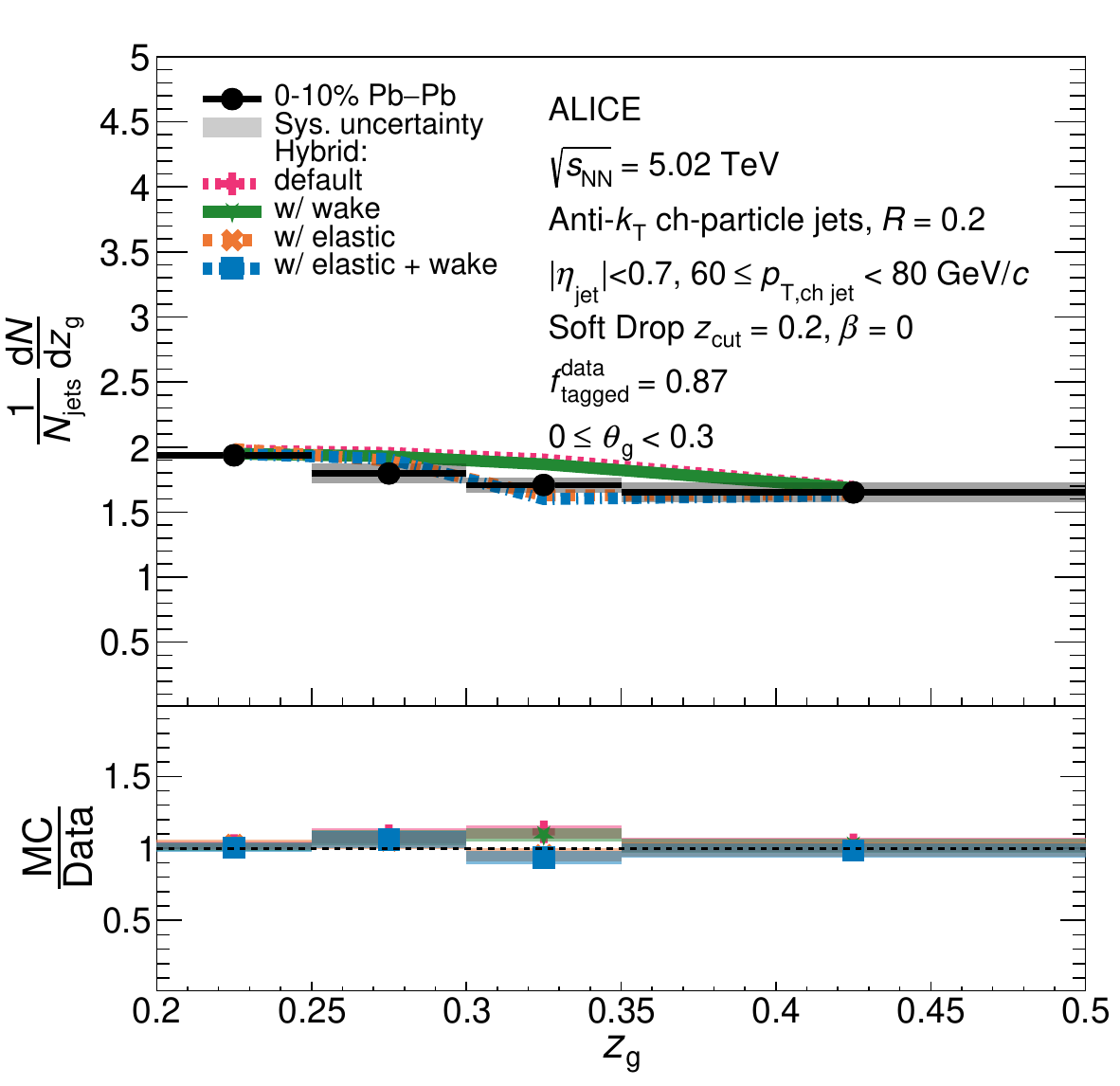}\hfill
  \includegraphics[width=.5\textwidth]{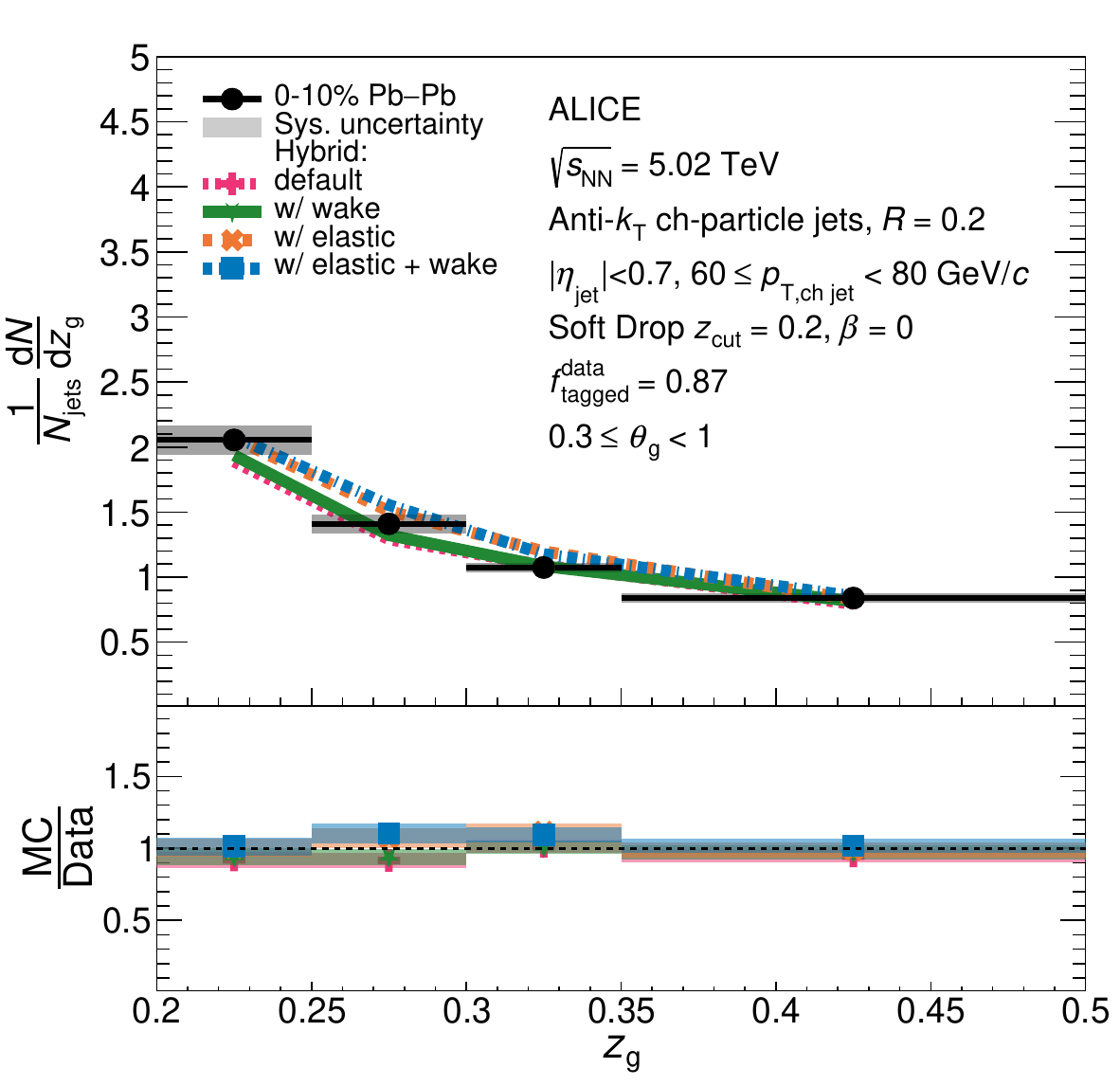}\hfill  
  \caption{The distributions of \zg in \PbPb collisions at \fivenn for jets with $60 \leq $ \ptjetch $< 80$ \GeVc for selections of small and large \rg. The left panels show the \zg distribution for $0.0 \leq \rg < 0.3$, and the right panels show \zg for $0.3 \leq \rg < 1.0$. Top panels compare the \JEWEL and \JETSCAPE models and bottom panels present the different \Hybrid tunes.}
  \label{fig:PbPbZgResults}
\end{figure}

Figure ~\ref{fig:PbPbZgResults} shows the \zg distribution in \PbPb collisions for selections of narrow and wide jets.
The selections for small and large \rg are the same as the selection for \pp collisions.
For narrow jets, \zg has a relatively flat distribution, while for wide jets the \zg distribution shows more of the characteristic $1/\zg$ behaviour.
This is reproduced by all the model predictions, which all describe the shape of the \zg distributions for both narrow and wide jets fairly well.
An overall shift in the measured yield of the \JEWEL distributions is observed, caused by the selection on \rg, where \JEWEL overestimates the jet narrowing effect.
For the \zg distributions presented here, \JEWEL predictions show no sensitivity to medium recoils, and the \Hybrid model shows no sensitivity to elastic scatterings or the medium wake. The latter suggesting that \zg in this kinematic regime does not discriminate between these model variants.

\begin{figure}[!h]
\centering
  \includegraphics[width=.5\textwidth]{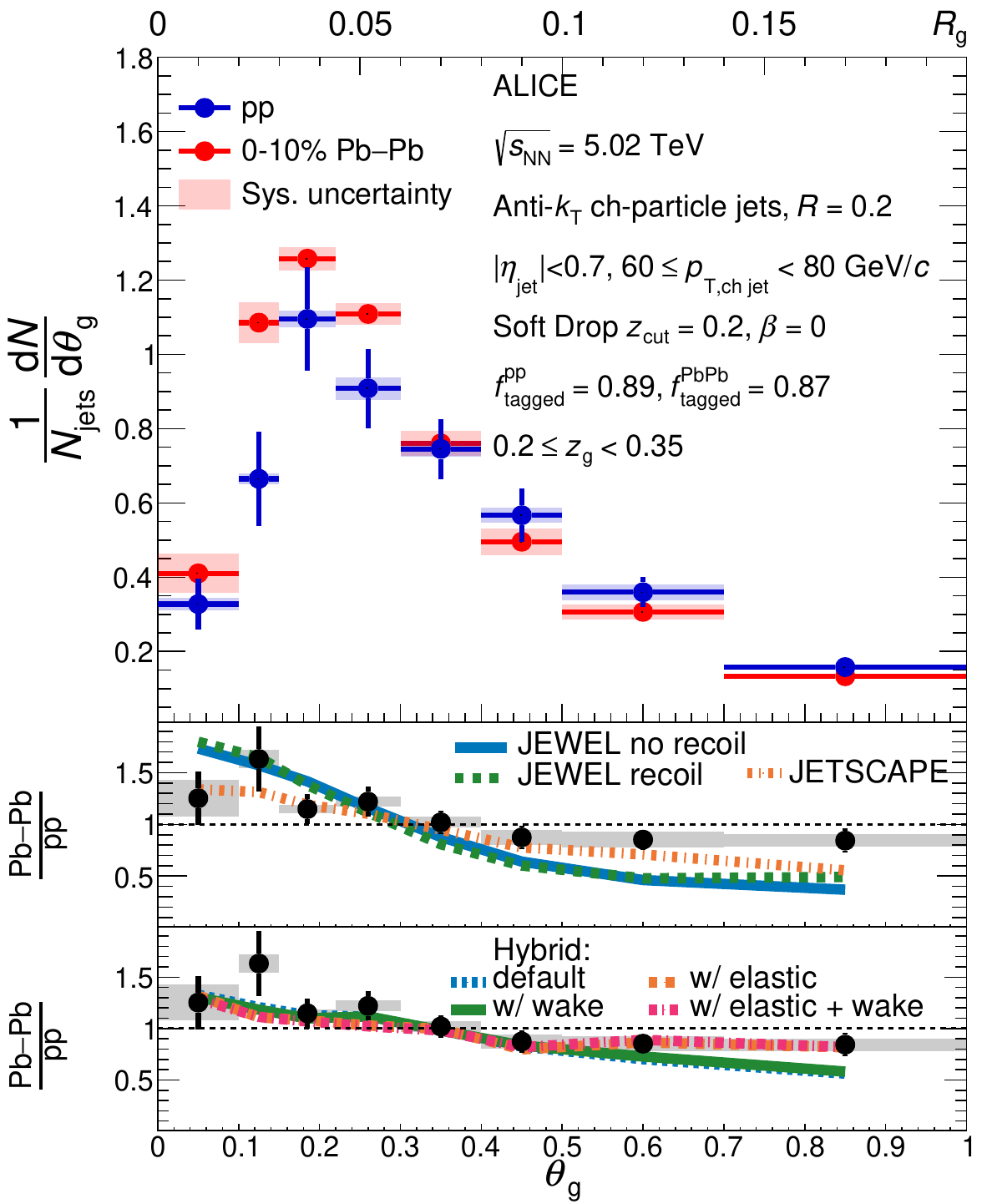}\hfill
  \includegraphics[width=.5\textwidth]{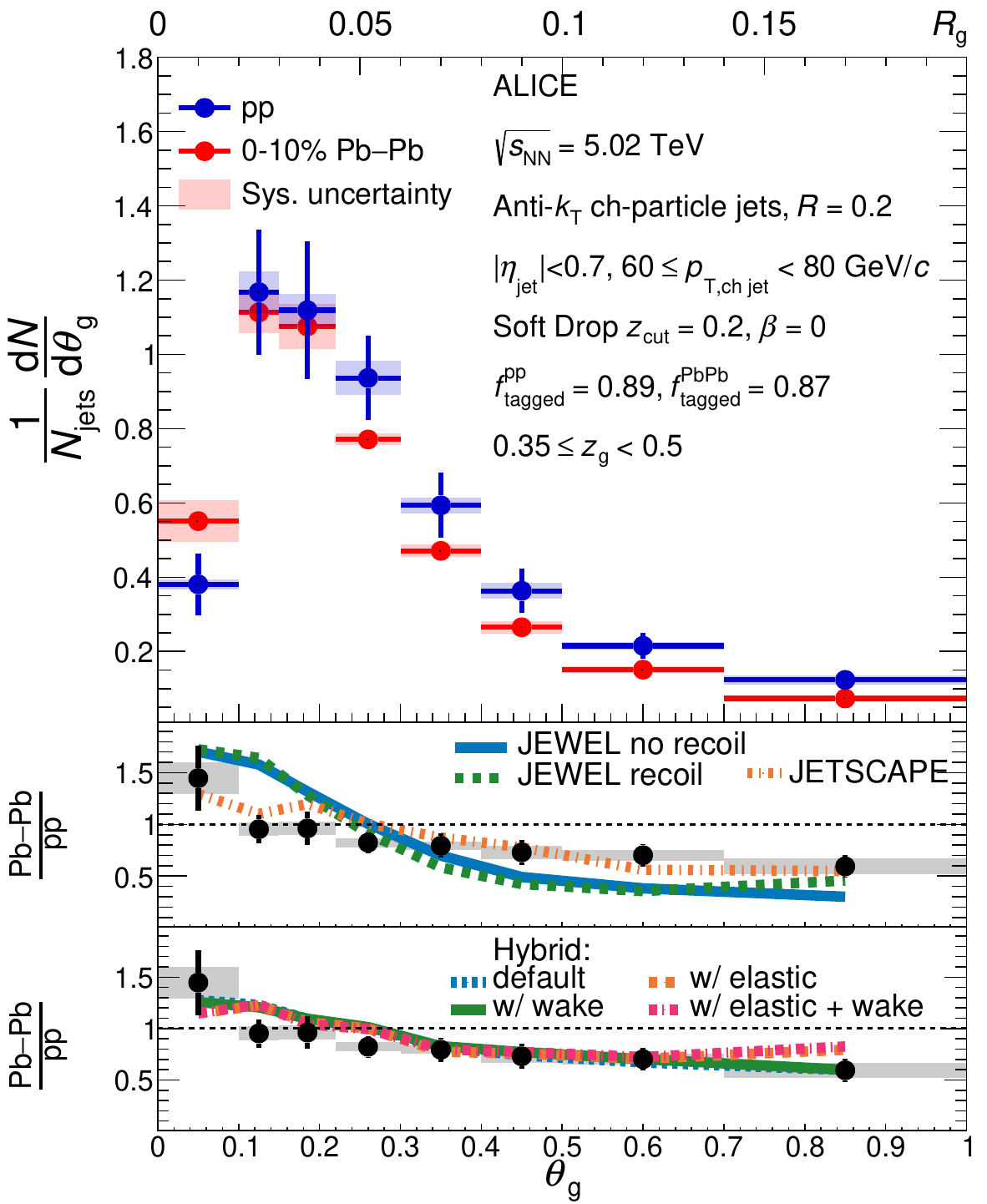}\hfill
  \caption{The distributions of \rg in \pp and \PbPb collisions at \fivenn for jets with $60 \leq $ \ptjetch $< 80$ \GeVc for selections of small and large \zg. The left panel shows the \rg distribution for $0.2 \leq \zg < 0.35$, and the right panel shows \rg for $0.35 \leq \zg < 0.5$. The bottom panels show the ratio of the measurement in \PbPb collisions to the measurement in \pp collisions.}
  \label{fig:ppPbPbRgResults}
\end{figure}

\clearpage
Figure~\ref{fig:ppPbPbRgResults} shows the \rg measurement in \pp and \PbPb collisions for unbalanced and balanced jets. The balanced jet population is observed to be narrower than the unbalanced jet population for \pp and \PbPb collisions. 
The bottom panels show the ratio of the measurements in \PbPb to \pp collisions, together with model predictions for \JEWEL, \JETSCAPE, and the \Hybrid model. 
In this ratio, jets measured in \PbPb collisions are observed to be narrower than in pp collisions. This effect is more pronounced for balanced jets. Wide jets are less likely to remain in the sample since they are expected to lose more energy, known as the survival bias. Therefore a narrowing of the distribution should be expected and it is observed to be of similar magnitude for unbalanced and balanced jets. 
\JEWEL tends to overestimate the jet narrowing effect for both low \zg and high \zg selections, and the jet narrowing shows little sensitivity to medium recoils.
The jet narrowing effect in \JETSCAPE describes the data well, but overestimates the depletion of jets at large \rg in the unbalanced jet sample.
From the \Hybrid model predictions, no sensitivity to the medium wake is seen. 
This might be because of the small jet radius or aggressive grooming of the jets that are measured.
At large \rg the \Hybrid model shows the effect of elastic scatterings, corresponding to a relative enhancement of the yield at large \rg.
In this large \rg region, the unbalanced subjet data favours the \Hybrid model with elastic scatterings, while the balanced subjet data prefers the model without elastic scatterings.

\begin{figure}[!h]
\centering
  \includegraphics[width=.5\textwidth]{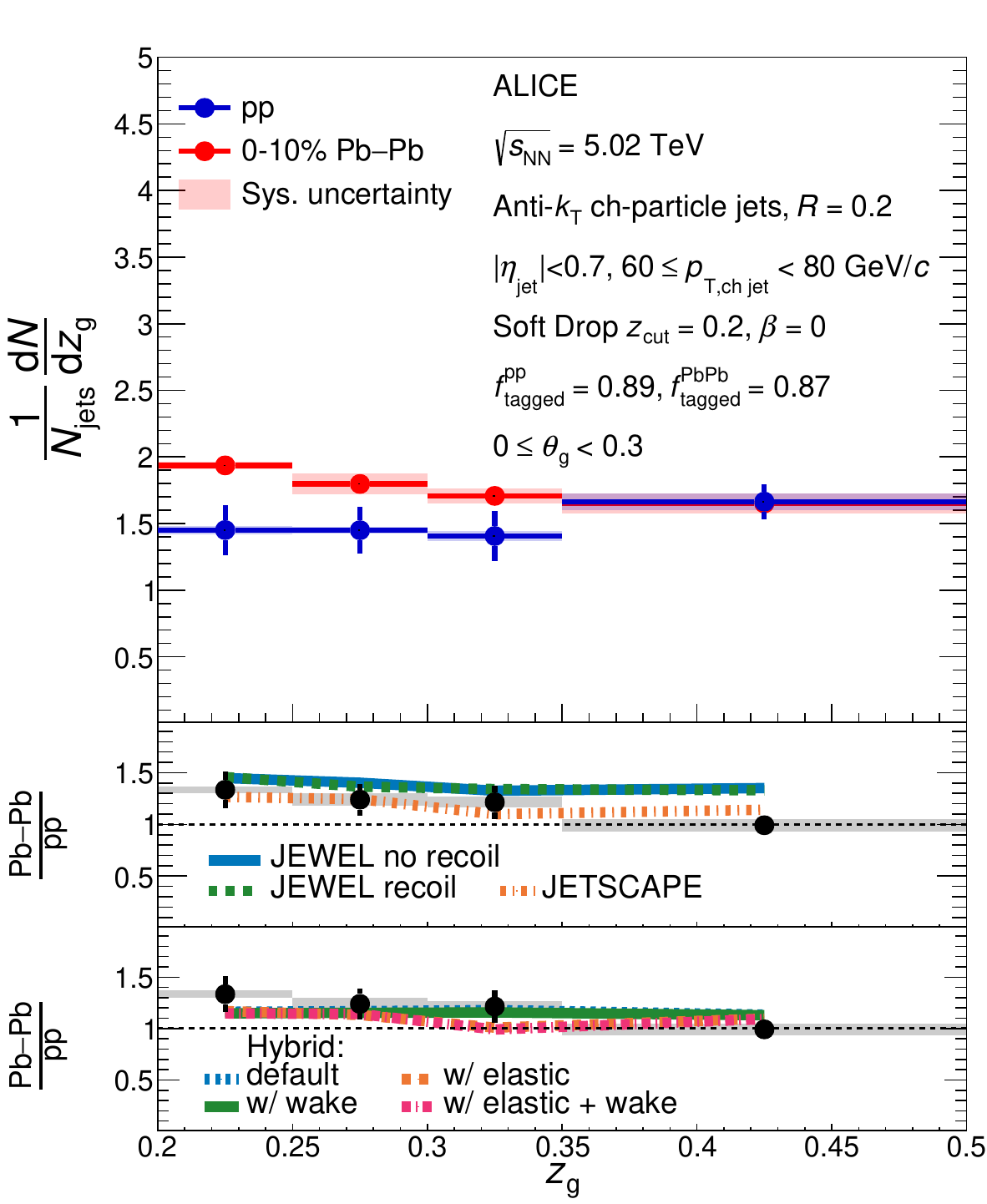}\hfill
  \includegraphics[width=.5\textwidth]{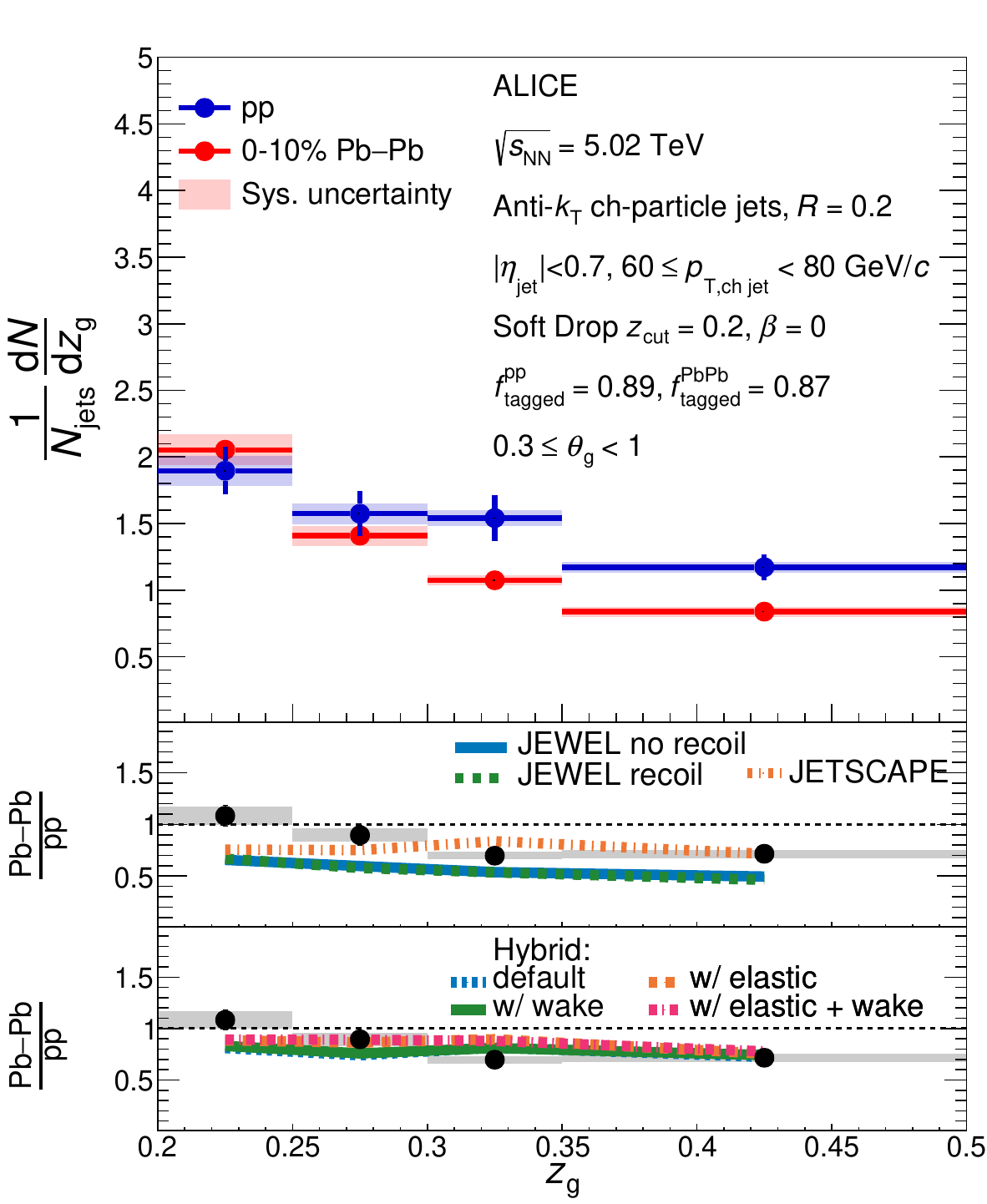}\hfill
  \caption{The measurement of \zg in \pp and \PbPb collisions at \fivenn for jets with $60 \leq $ \ptjetch $< 80$ \GeVc for selections of small and large \rg. The left panel shows the \zg distribution for $0.0 \leq \rg < 0.3$, and the right panel shows \zg for $0.3 \leq \rg < 1.0$. The bottom panels show the ratio of the measurement in \PbPb collisions to the measurement in \pp collisions.}
  \label{fig:ppPbPbZgResults}
\end{figure}

Figure~\ref{fig:ppPbPbZgResults} shows the measurement of \zg for \pp and \PbPb collisions for narrow and wide jets.
The bottom panels show the ratio of the measurement in \PbPb collisions to the measurement in \pp collisions, together with model predictions.
By selecting narrow or wide jets, the sensitivity to jet-quenching effects can be varied.
The selection of narrow jets does not show any significant modification of \zg within statistical and systematic uncertainties, as can be seen from the relatively flat ratio of the \PbPb and \pp collision measurements in the left panel of Fig.~\ref{fig:ppPbPbZgResults}.
The \zg measurement for wide jets shows a small slope in the ratio, indicating that wide jets are less balanced in \PbPb collisions.
This ratio can be fit with a constant value corresponding to the ratio of the integral of the \PbPb and \pp distributions.
The obtained value is approximately $0.82$.
Taking into account the statistical error, the modification with respect to this line is 2.7$\sigma$.
Adding the systematic uncertainty in quadrature lowers the significance to 2.0$\sigma$, where the correlation of the systematic uncertainties was taken into account.

The significance of the measurement is mainly limited by the statistical error in the \pp collision measurement. Balanced jets, represented through data points at large \zg, drive the observed difference between the \PbPb and \pp results particularly for wider jets. All models show relatively flat \PbPb to \pp ratios, as observed in the data. The amount of jets that enter the ratios are defined by the \rg selection which, as shown in Fig.~\ref{fig:PbPbRgResults}, is much narrower for \JEWEL than in data. 
As a result the ratios from Fig.~\ref{fig:ppPbPbRgResults} and Fig.~\ref{fig:ppPbPbZgResults} at small \rg are above and at large \rg below unity.

\section{Conclusions}
The correlation between the Soft Drop groomed jet splitting angle \rg and momentum-sharing  fraction \zg is reported for jets with $60 \leq $ \ptjetch $< 80$ \GeVc ($\zcut = 0.2$, $\beta = 0$) in \pp and central \PbPb collisions.
A jet narrowing effect for unbalanced ($\zg < 0.35$) and balanced ($z_g \approx 0.5$) subjets in \PbPb collisions is observed.
This effect is overestimated in \JEWEL, while it is well described by \JETSCAPE.
The \Hybrid model predictions of \rg suggest that this measurement is sensitive to elastic scatterings.
For unbalanced subjets, the data favours the jet narrowing prediction in the \Hybrid model variant with elastic scatterings, while for balanced subjets the variant without elastic scatterings is preferred.
Due to the limited jet radius, the measurement is insensitive to the different implementations of medium response. 
There is a hint that splittings with a large opening angle are less balanced in transverse momentum in \PbPb collisions compared to \pp collisions, while no such modification is present for narrow subjets. This can be understood as a result of the fact that jets with a larger opening angle have more constituent partons which lose energy independently, leading to a larger energy loss for wide jets than for narrow jets. 
This effect, combined with the change of the \zg distribution with jet energy may affect the shape of the \zg distribution. Direct effects of the energy loss of subjets on the jet structure are also likely to contribute at some level.


\newenvironment{acknowledgement}{\relax}{\relax}
\begin{acknowledgement}
\section*{Acknowledgements}

The ALICE Collaboration would like to thank all its engineers and technicians for their invaluable contributions to the construction of the experiment and the CERN accelerator teams for the outstanding performance of the LHC complex.
The ALICE Collaboration gratefully acknowledges the resources and support provided by all Grid centres and the Worldwide LHC Computing Grid (WLCG) collaboration.
The ALICE Collaboration acknowledges the following funding agencies for their support in building and running the ALICE detector:
A. I. Alikhanyan National Science Laboratory (Yerevan Physics Institute) Foundation (ANSL), State Committee of Science and World Federation of Scientists (WFS), Armenia.
Austrian Academy of Sciences, Austrian Science Fund (FWF): [M 2467-N36] and Nationalstiftung f\"{u}r Forschung, Technologie und Entwicklung, Austria.
Ministry of Communications and High Technologies, National Nuclear Research Center, Azerbaijan.
Rede Nacional de Física de Altas Energias (Renafae), Financiadora de Estudos e Projetos (Finep), Funda\c{c}\~{a}o de Amparo \`{a} Pesquisa do Estado de S\~{a}o Paulo (FAPESP) and The Sao Paulo Research Foundation  (FAPESP), Brazil.
Bulgarian Ministry of Education and Science, within the National Roadmap for Research Infrastructures 2020-2027 (object CERN), Bulgaria.
Ministry of Education of China (MOEC) , Ministry of Science \& Technology of China (MSTC) and National Natural Science Foundation of China (NSFC), China.
Ministry of Science and Education and Croatian Science Foundation, Croatia.
Centro de Aplicaciones Tecnol\'{o}gicas y Desarrollo Nuclear (CEADEN), Cubaenerg\'{\i}a, Cuba.
Ministry of Education, Youth and Sports of the Czech Republic, Czech Republic.
The Danish Council for Independent Research | Natural Sciences, the VILLUM FONDEN and Danish National Research Foundation (DNRF), Denmark.
Helsinki Institute of Physics (HIP), Finland.
Commissariat \`{a} l'Energie Atomique (CEA) and Institut National de Physique Nucl\'{e}aire et de Physique des Particules (IN2P3) and Centre National de la Recherche Scientifique (CNRS), France.
Bundesministerium f\"{u}r Forschung, Technologie und Raumfahrt (BMFTR) and GSI Helmholtzzentrum f\"{u}r Schwerionenforschung GmbH, Germany.
National Research, Development and Innovation Office, Hungary.
Department of Atomic Energy Government of India (DAE), Department of Science and Technology, Government of India (DST), University Grants Commission, Government of India (UGC) and Council of Scientific and Industrial Research (CSIR), India.
National Research and Innovation Agency - BRIN, Indonesia.
Istituto Nazionale di Fisica Nucleare (INFN), Italy.
Japanese Ministry of Education, Culture, Sports, Science and Technology (MEXT) and Japan Society for the Promotion of Science (JSPS) KAKENHI, Japan.
Consejo Nacional de Ciencia (CONACYT) y Tecnolog\'{i}a, through Fondo de Cooperaci\'{o}n Internacional en Ciencia y Tecnolog\'{i}a (FONCICYT) and Direcci\'{o}n General de Asuntos del Personal Academico (DGAPA), Mexico.
Nederlandse Organisatie voor Wetenschappelijk Onderzoek (NWO), Netherlands.
The Research Council of Norway, Norway.
Pontificia Universidad Cat\'{o}lica del Per\'{u}, Peru.
Ministry of Science and Higher Education, National Science Centre and WUT ID-UB, Poland.
Korea Institute of Science and Technology Information and National Research Foundation of Korea (NRF), Republic of Korea.
Ministry of Education and Scientific Research, Institute of Atomic Physics, Ministry of Research and Innovation and Institute of Atomic Physics and Universitatea Nationala de Stiinta si Tehnologie Politehnica Bucuresti, Romania.
Ministerstvo skolstva, vyskumu, vyvoja a mladeze SR, Slovakia.
National Research Foundation of South Africa, South Africa.
Swedish Research Council (VR) and Knut \& Alice Wallenberg Foundation (KAW), Sweden.
European Organization for Nuclear Research, Switzerland.
Suranaree University of Technology (SUT), National Science and Technology Development Agency (NSTDA) and National Science, Research and Innovation Fund (NSRF via PMU-B B05F650021), Thailand.
Turkish Energy, Nuclear and Mineral Research Agency (TENMAK), Turkey.
National Academy of  Sciences of Ukraine, Ukraine.
Science and Technology Facilities Council (STFC), United Kingdom.
National Science Foundation of the United States of America (NSF) and United States Department of Energy, Office of Nuclear Physics (DOE NP), United States of America.
In addition, individual groups or members have received support from:
Czech Science Foundation (grant no. 23-07499S), Czech Republic.
FORTE project, reg.\ no.\ CZ.02.01.01/00/22\_008/0004632, Czech Republic, co-funded by the European Union, Czech Republic.
European Research Council (grant no. 101220549), European Union.
Deutsche Forschungs Gemeinschaft (DFG, German Research Foundation) ``Neutrinos and Dark Matter in Astro- and Particle Physics'' (grant no. SFB 1258), Germany.
CONVECS project, CUP C97H23001700002 FESR 2021-2027 program, Italy.

\end{acknowledgement}

\bibliographystyle{utphys}   
\bibliography{bibliography}

\newpage
\appendix
\section{Appendix}
\begin{figure}[!h]
\centering
  \includegraphics[width=.5\textwidth]{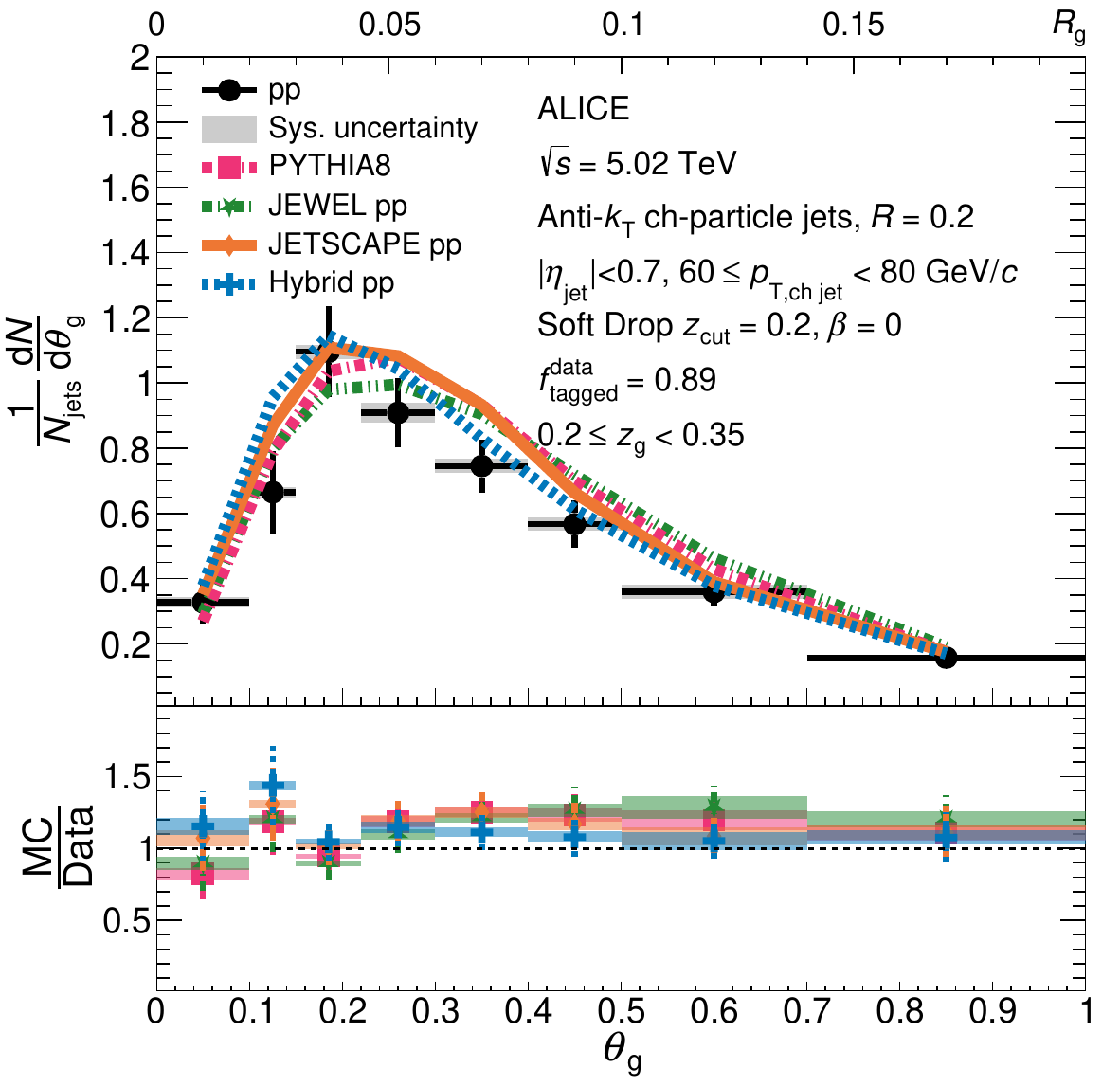}\hfill
  \includegraphics[width=.5\textwidth]{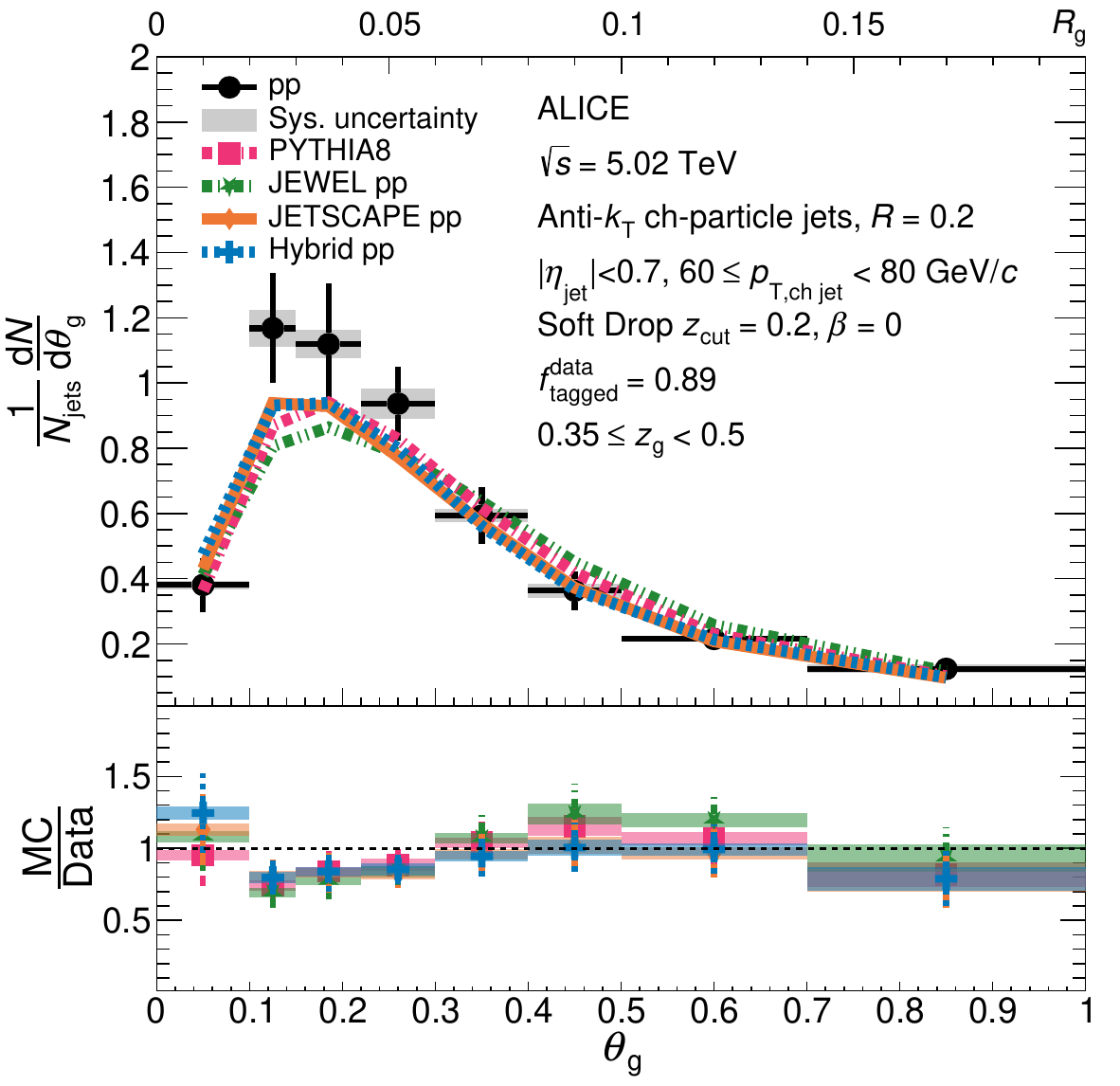}\hfill
  \caption{The distributions of \rg in \pp collisions, compared to jet-quenching \pp baselines, at \five for jets with $60 \leq $ \ptjetch $< 80$ \GeVc, for selections of small and large \zg. The left panel shows the \rg distribution for $0.2 \leq \zg < 0.35$, and the right panel shows \rg for $0.35 \leq \zg < 0.5$.}
  \label{fig:baselineRgResults}
\end{figure}

\begin{figure}[!h]
\centering
  \includegraphics[width=.5\textwidth]{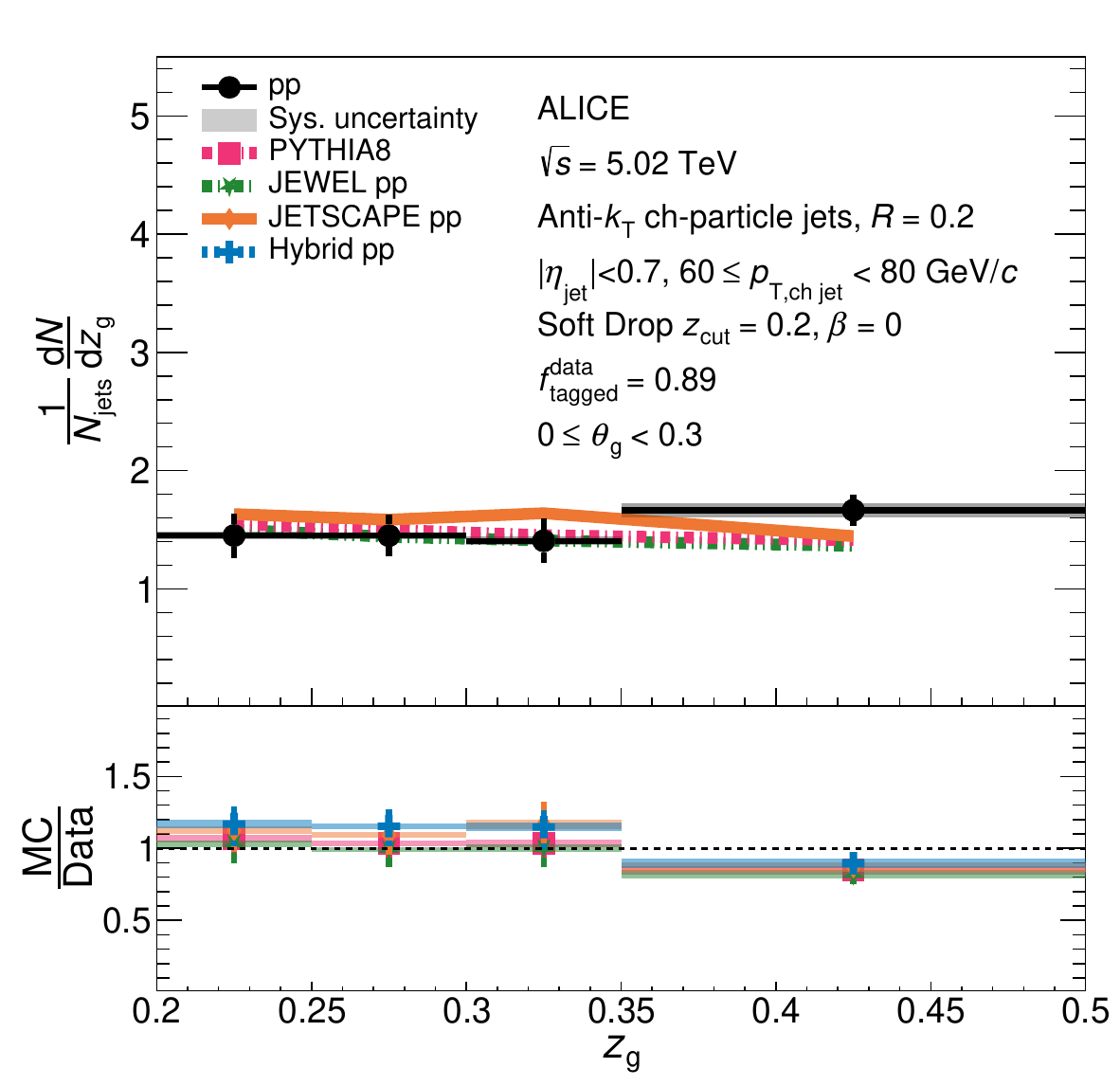}\hfill
  \includegraphics[width=.5\textwidth]{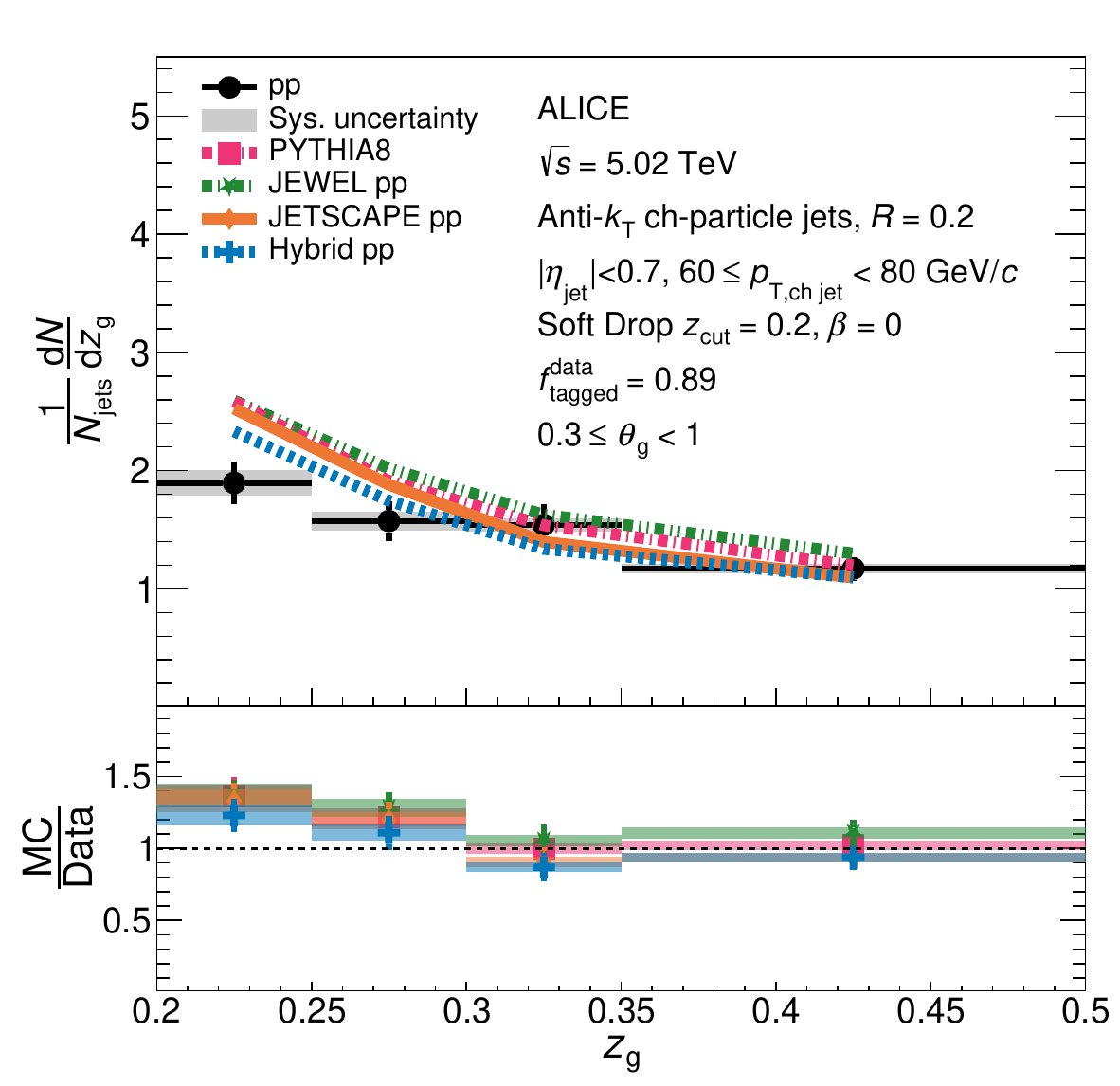}\hfill
  \caption{The distributions of \zg in \pp collisions, compared to jet-quenching \pp baselines, at \five for jets with $60 \leq $ \ptjetch $< 80$ \GeVc, for selections of small and large \rg. The left panel shows the \zg distribution for $0.0 \leq \rg < 0.3$, and the right panel shows \zg for $0.3 \leq \rg < 1.0$.}
  \label{fig:baselineZgResults}
\end{figure}

Figure~\ref{fig:baselineRgResults} shows the distribution of \rg in \pp collisions for unbalanced and balanced jets, together with \pp collision model predictions from \PYTHIA and the vacuum mode of the \JEWEL, \JETSCAPE and \Hybrid model jet-quenching event generators.
Similarly, Fig.~\ref{fig:baselineZgResults} shows the \zg distribution in \pp collisions for selections of narrow and wide jets together with \PYTHIA and jet-quenching model \pp collision predictions.

\begin{figure}[!h]
\centering
  \includegraphics[width=.5\textwidth]{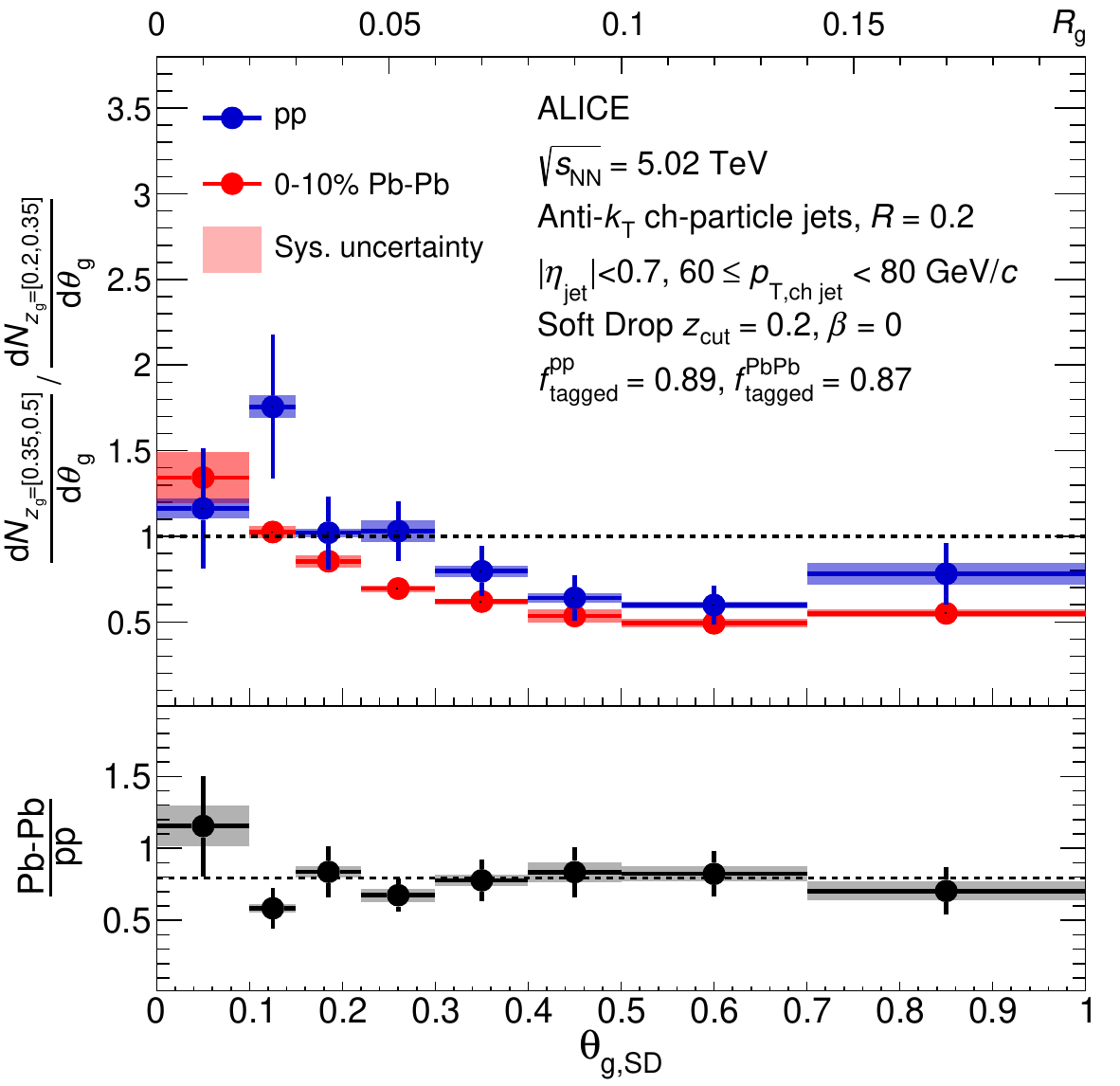}\hfill
  \caption{The ratio of the \rg distributions for the low and high \zg selections for \pp and \PbPb collisions at \fivenn for jets with $60 \leq $ \ptjetch $< 80$ \GeVc. The ratios shows that in \PbPb collisions the high \zg selection has significantly narrower jets compared to \pp collisions. The dashed line in the bottom panel represents the ratio of the integral of the \PbPb collision distribution to the integral of the \pp collision distribution.}
  \label{fig:rgRatioRatio}
\end{figure}

In Fig.~\ref{fig:rgRatioRatio} the measurement of the double ratio of \rg for balanced and unbalanced subjets in \pp and \PbPb collisions is shown.
By comparing \rg for balanced and unbalanced subjets in the same collision system, the systematic uncertainty is partially cancelled.
The top panel shows the ratio of \rg in balanced and unbalanced subjets for \pp and \PbPb collisions.
In both collision systems, balanced subjets are more narrow. 
The bottom panel shows the double ratio of \PbPb and \pp collisions.
The integrals of the distributions for \PbPb and \pp collisions are not equal because of the modification of \zg in \PbPb jets.
Because the ratio is constant, the effect of more balanced jets being more narrow is not stronger in \PbPb than in \pp collisions.
This indicates that the jet narrowing effect in \PbPb collisions is independent of \zg. In summary, no difference in the jet narrowing effect is found for unbalanced and balanced subjets other than the modification of the yield of the \zg selections.

\newpage

\section{The ALICE Collaboration}
\label{app:collab}
\begin{flushleft} 
\small

D.A.H.~Abdallah\,\orcidlink{0000-0003-4768-2718}\,$^{\rm 134}$, 
I.J.~Abualrob\,\orcidlink{0009-0005-3519-5631}\,$^{\rm 111}$, 
S.~Acharya\,\orcidlink{0000-0002-9213-5329}\,$^{\rm 49}$, 
K.~Agarwal\,\orcidlink{0000-0001-5781-3393}\,$^{\rm II,}$$^{\rm 23}$, 
G.~Aglieri Rinella\,\orcidlink{0000-0002-9611-3696}\,$^{\rm 32}$, 
L.~Aglietta\,\orcidlink{0009-0003-0763-6802}\,$^{\rm 24}$, 
N.~Agrawal\,\orcidlink{0000-0003-0348-9836}\,$^{\rm 25}$, 
Z.~Ahammed\,\orcidlink{0000-0001-5241-7412}\,$^{\rm 132}$, 
S.~Ahmad\,\orcidlink{0000-0003-0497-5705}\,$^{\rm 15}$, 
Z.~Akbar\,\orcidlink{0000-0002-5373-6121}\,$^{\rm 79}$
% ...

\section*{Affiliation Notes}

$^{\rm I}$ Deceased\\
$^{\rm II}$ Also at: INFN Trieste, Trieste, Italy\\
$^{\rm III}$ Also at: Fondazione Bruno Kessler (FBK), Trento, Italy\\
$^{\rm IV}$ Also at: Czech Technical University in Prague, Prague, Czech Republic\\
$^{\rm V}$ Also at: Dipartimento DET del Politecnico di Torino, Turin, Italy\\
$^{\rm VI}$ Also at: University College of Dublin, Dublin, Ireland\\
$^{\rm VII}$ Also at: Department of Applied Physics, Aligarh Muslim University, Aligarh, India\\
$^{\rm VIII}$ Also at: Institute of Theoretical Physics, University of Wroclaw, Wroclaw, Poland\\
$^{\rm IX}$ Also at: Facultad de Ciencias, Universidad Nacional Aut\'{o}noma de M\'{e}xico, Mexico City, Mexico\\
$^{\rm X}$ Also at: Lovely Professional University, Punjab, India
\\

\section*{Collaboration Institutes}

$^{1}$ A.I. Alikhanyan National Science Laboratory (Yerevan Physics Institute) Foundation, Yerevan, Armenia\\
$^{2}$ AGH University of Krakow, Cracow, Poland\\
$^{3}$ Bogolyubov Institute for Theoretical Physics, National Academy of Sciences of Ukraine, Kyiv, Ukraine\\
$^{4}$ Bose Institute, Department of Physics  and Centre for Astroparticle Physics and Space Science (CAPSS), Kolkata, India\\
$^{5}$ California Polytechnic State University, San Luis Obispo, California, United States\\
$^{6}$ Central China Normal University, Wuhan, China\\
$^{7}$ Centro de Aplicaciones Tecnol\'{o}gicas y Desarrollo Nuclear (CEADEN), Havana, Cuba\\
$^{8}$ Centro de Investigaci\'{o}n y de Estudios Avanzados (CINVESTAV), Mexico City and M\'{e}rida, Mexico\\
$^{9}$ Chicago State University, Chicago, Illinois, United States\\
$^{10}$ China Nuclear Data Center, China Institute of Atomic Energy, Beijing, China\\
$^{11}$ China University of Geosciences, Wuhan, China\\
$^{12}$ Chungbuk National University, Cheongju, Republic of Korea\\
$^{13}$ Comenius University Bratislava, Faculty of Mathematics, Physics and Informatics, Bratislava, Slovak Republic\\
$^{14}$ Creighton University, Omaha, Nebraska, United States\\
$^{15}$ Department of Physics, Aligarh Muslim University, Aligarh, India\\
$^{16}$ Department of Physics, Pusan National University, Pusan, Republic of Korea\\
$^{17}$ Department of Physics, Sejong University, Seoul, Republic of Korea\\
$^{18}$ Department of Physics, University of California, Berkeley, California, United States\\
$^{19}$ Department of Physics, University of Oslo, Oslo, Norway\\
$^{20}$ Department of Physics and Technology, University of Bergen, Bergen, Norway\\
$^{21}$ Dipartimento di Fisica, Universit\`{a} di Pavia, Pavia, Italy\\
$^{22}$ Dipartimento di Fisica dell'Universit\`{a} and Sezione INFN, Cagliari, Italy\\
$^{23}$ Dipartimento di Fisica dell'Universit\`{a} and Sezione INFN, Trieste, Italy\\
$^{24}$ Dipartimento di Fisica dell'Universit\`{a} and Sezione INFN, Turin, Italy\\
$^{25}$ Dipartimento di Fisica e Astronomia dell'Universit\`{a} and Sezione INFN, Bologna, Italy\\
$^{26}$ Dipartimento di Fisica e Astronomia dell'Universit\`{a} and Sezione INFN, Catania, Italy\\
$^{27}$ Dipartimento di Fisica e Astronomia dell'Universit\`{a} and Sezione INFN, Padova, Italy\\
$^{28}$ Dipartimento di Fisica `E.R.~Caianiello' dell'Universit\`{a} and Gruppo Collegato INFN, Salerno, Italy\\
$^{29}$ Dipartimento DISAT del Politecnico and Sezione INFN, Turin, Italy\\
$^{30}$ Dipartimento di Scienze MIFT, Universit\`{a} di Messina, Messina, Italy\\
$^{31}$ Dipartimento Interateneo di Fisica `M.~Merlin' and Sezione INFN, Bari, Italy\\
$^{32}$ European Organization for Nuclear Research (CERN), Geneva, Switzerland\\
$^{33}$ Faculty of Electrical Engineering, Mechanical Engineering and Naval Architecture, University of Split, Split, Croatia\\
$^{34}$ Faculty of Nuclear Sciences and Physical Engineering, Czech Technical University in Prague, Prague, Czech Republic\\
$^{35}$ Faculty of Physics, Sofia University, Sofia, Bulgaria\\
$^{36}$ Faculty of Science, P.J.~\v{S}af\'{a}rik University, Ko\v{s}ice, Slovak Republic\\
$^{37}$ Faculty of Technology, Environmental and Social Sciences, Bergen, Norway\\
$^{38}$ Frankfurt Institute for Advanced Studies, Johann Wolfgang Goethe-Universit\"{a}t Frankfurt, Frankfurt, Germany\\
$^{39}$ Fudan University, Shanghai, China\\
$^{40}$ Gauhati University, Department of Physics, Guwahati, India\\
$^{41}$ Helmholtz-Institut f\"{u}r Strahlen- und Kernphysik, Rheinische Friedrich-Wilhelms-Universit\"{a}t Bonn, Bonn, Germany\\
$^{42}$ Helsinki Institute of Physics (HIP), Helsinki, Finland\\
$^{43}$ High Energy Physics Group,  Universidad Aut\'{o}noma de Puebla, Puebla, Mexico\\
$^{44}$ Horia Hulubei National Institute of Physics and Nuclear Engineering, Bucharest, Romania\\
$^{45}$ HUN-REN Wigner Research Centre for Physics, Budapest, Hungary\\
$^{46}$ Indian Institute of Technology Bombay (IIT), Mumbai, India\\
$^{47}$ Indian Institute of Technology Indore, Indore, India\\
$^{48}$ INFN, Laboratori Nazionali di Frascati, Frascati, Italy\\
$^{49}$ INFN, Sezione di Bari, Bari, Italy\\
$^{50}$ INFN, Sezione di Bologna, Bologna, Italy\\
$^{51}$ INFN, Sezione di Cagliari, Cagliari, Italy\\
$^{52}$ INFN, Sezione di Catania, Catania, Italy\\
$^{53}$ INFN, Sezione di Padova, Padova, Italy\\
$^{54}$ INFN, Sezione di Pavia, Pavia, Italy\\
$^{55}$ INFN, Sezione di Torino, Turin, Italy\\
$^{56}$ INFN, Sezione di Trieste, Trieste, Italy\\
$^{57}$ Inha University, Incheon, Republic of Korea\\
$^{58}$ Institute for Gravitational and Subatomic Physics (GRASP), Utrecht University/Nikhef, Utrecht, Netherlands\\
$^{59}$ Institute of Experimental Physics, Slovak Academy of Sciences, Ko\v{s}ice, Slovak Republic\\
$^{60}$ Institute of Physics, Homi Bhabha National Institute, Bhubaneswar, India\\
$^{61}$ Institute of Physics of the Czech Academy of Sciences, Prague, Czech Republic\\
$^{62}$ Institute of Space Science (ISS), Bucharest, Romania\\
$^{63}$ Institut f\"{u}r Kernphysik, Johann Wolfgang Goethe-Universit\"{a}t Frankfurt, Frankfurt, Germany\\
$^{64}$ Instituto de Ciencias Nucleares, Universidad Nacional Aut\'{o}noma de M\'{e}xico, Mexico City, Mexico\\
$^{65}$ Instituto de F\'{i}sica, Universidade Federal do Rio Grande do Sul (UFRGS), Porto Alegre, Brazil\\
$^{66}$ Instituto de F\'{\i}sica, Universidad Nacional Aut\'{o}noma de M\'{e}xico, Mexico City, Mexico\\
$^{67}$ iThemba LABS, National Research Foundation, Somerset West, South Africa\\
$^{68}$ Jeonbuk National University, Jeonju, Republic of Korea\\
$^{69}$ Korea Institute of Science and Technology Information, Daejeon, Republic of Korea\\
$^{70}$ Laboratoire de Physique Subatomique et de Cosmologie, Universit\'{e} Grenoble-Alpes, CNRS-IN2P3, Grenoble, France\\
$^{71}$ Lawrence Berkeley National Laboratory, Berkeley, California, United States\\
$^{72}$ Lund University Department of Physics, Division of Particle Physics, Lund, Sweden\\
$^{73}$ Marietta Blau Institute, Vienna, Austria\\
$^{74}$ Nagasaki Institute of Applied Science, Nagasaki, Japan\\
$^{75}$ Nara Women{'}s University (NWU), Nara, Japan\\
$^{76}$ National Centre for Nuclear Research, Warsaw, Poland\\
$^{77}$ National Institute of Science Education and Research, Homi Bhabha National Institute, Jatni, India\\
$^{78}$ National Nuclear Research Center, Baku, Azerbaijan\\
$^{79}$ National Research and Innovation Agency - BRIN, Jakarta, Indonesia\\
$^{80}$ Niels Bohr Institute, University of Copenhagen, Copenhagen, Denmark\\
$^{81}$ Nikhef, National institute for subatomic physics, Amsterdam, Netherlands\\
$^{82}$ Nuclear Physics Group, STFC Daresbury Laboratory, Daresbury, United Kingdom\\
$^{83}$ Nuclear Physics Institute of the Czech Academy of Sciences, Husinec-\v{R}e\v{z}, Czech Republic\\
$^{84}$ Oak Ridge National Laboratory, Oak Ridge, Tennessee, United States\\
$^{85}$ Ohio State University, Columbus, Ohio, United States\\
$^{86}$ Physics Department, Panjab University, Chandigarh, India\\
$^{87}$ Physics Department, University of Jammu, Jammu, India\\
$^{88}$ Physics Program and International Institute for Sustainability with Knotted Chiral Meta Matter (WPI-SKCM$^{2}$), Hiroshima University, Hiroshima, Japan\\
$^{89}$ Physikalisches Institut, Eberhard-Karls-Universit\"{a}t T\"{u}bingen, T\"{u}bingen, Germany\\
$^{90}$ Physikalisches Institut, Ruprecht-Karls-Universit\"{a}t Heidelberg, Heidelberg, Germany\\
$^{91}$ Physik Department, Technische Universit\"{a}t M\"{u}nchen, Munich, Germany\\
$^{92}$ Politecnico di Bari and Sezione INFN, Bari, Italy\\
$^{93}$ Research Division and ExtreMe Matter Institute EMMI, GSI Helmholtzzentrum f\"ur Schwerionenforschung GmbH, Darmstadt, Germany\\
$^{94}$ Saga University, Saga, Japan\\
$^{95}$ Saha Institute of Nuclear Physics, Homi Bhabha National Institute, Kolkata, India\\
$^{96}$ School of Physics and Astronomy, University of Birmingham, Birmingham, United Kingdom\\
$^{97}$ Secci\'{o}n F\'{\i}sica, Departamento de Ciencias, Pontificia Universidad Cat\'{o}lica del Per\'{u}, Lima, Peru\\
$^{98}$ SUBATECH, IMT Atlantique, Nantes Universit\'{e}, CNRS-IN2P3, Nantes, France\\
$^{99}$ Sungkyunkwan University, Suwon City, Republic of Korea\\
$^{100}$ Suranaree University of Technology, Nakhon Ratchasima, Thailand\\
$^{101}$ Technical University of Ko\v{s}ice, Ko\v{s}ice, Slovak Republic\\
$^{102}$ The Henryk Niewodniczanski Institute of Nuclear Physics, Polish Academy of Sciences, Cracow, Poland\\
$^{103}$ The University of Texas at Austin, Austin, Texas, United States\\
$^{104}$ Universidad Aut\'{o}noma de Sinaloa, Culiac\'{a}n, Mexico\\
$^{105}$ Universidade de S\~{a}o Paulo (USP), S\~{a}o Paulo, Brazil\\
$^{106}$ Universidade Estadual de Campinas (UNICAMP), Campinas, Brazil\\
$^{107}$ Universidade Federal do ABC, Santo Andre, Brazil\\
$^{108}$ Universitatea Nationala de Stiinta si Tehnologie Politehnica Bucuresti, Bucharest, Romania\\
$^{109}$ University of Cape Town, Cape Town, South Africa\\
$^{110}$ University of Derby, Derby, United Kingdom\\
$^{111}$ University of Houston, Houston, Texas, United States\\
$^{112}$ University of Jyv\"{a}skyl\"{a}, Jyv\"{a}skyl\"{a}, Finland\\
$^{113}$ University of Kansas, Lawrence, Kansas, United States\\
$^{114}$ University of Liverpool, Liverpool, United Kingdom\\
$^{115}$ University of Science and Technology of China, Hefei, China\\
$^{116}$ University of Silesia in Katowice, Katowice, Poland\\
$^{117}$ University of South-Eastern Norway, Kongsberg, Norway\\
$^{118}$ University of Tennessee, Knoxville, Tennessee, United States\\
$^{119}$ University of the Witwatersrand, Johannesburg, South Africa\\
$^{120}$ University of Tokyo, Tokyo, Japan\\
$^{121}$ University of Tsukuba, Tsukuba, Japan\\
$^{122}$ University of Zagreb Faculty of Science, Department of Physics, Zagreb, Croatia\\
$^{123}$ Universit\"{a}t M\"{u}nster, Institut f\"{u}r Kernphysik, M\"{u}nster, Germany\\
$^{124}$ Universit\'{e} Clermont Auvergne, CNRS/IN2P3, LPC, Clermont-Ferrand, France\\
$^{125}$ Universit\'{e} de Lyon, CNRS/IN2P3, Institut de Physique des 2 Infinis de Lyon, Lyon, France\\
$^{126}$ Universit\'{e} de Strasbourg, CNRS, IPHC UMR 7178, F-67000 Strasbourg, France, Strasbourg, France\\
$^{127}$ Universit\'{e} Paris-Saclay, Centre d'Etudes de Saclay (CEA), IRFU, D\'{e}partment de Physique Nucl\'{e}aire (DPhN), Saclay, France\\
$^{128}$ Universit\'{e}  Paris-Saclay, CNRS/IN2P3, IJCLab, Orsay, France\\
$^{129}$ Universit\`{a} degli Studi di Foggia, Foggia, Italy\\
$^{130}$ Universit\`{a} del Piemonte Orientale, Vercelli, Italy\\
$^{131}$ Universit\`{a} di Brescia, Brescia, Italy\\
$^{132}$ Variable Energy Cyclotron Centre, Homi Bhabha National Institute, Kolkata, India\\
$^{133}$ Warsaw University of Technology, Warsaw, Poland\\
$^{134}$ Wayne State University, Detroit, Michigan, United States\\
$^{135}$ Yale University, New Haven, Connecticut, United States\\
$^{136}$ Yildiz Technical University, Istanbul, Turkey\\
$^{137}$ Yonsei University, Seoul, Republic of Korea\\
$^{138}$ Affiliated with an institute formerly covered by a cooperation agreement with CERN\\
$^{139}$ Affiliated with an international laboratory covered by a cooperation agreement with CERN.\\

\end{flushleft} 
  
\end{document}